%% file: ms.tex
\documentclass[%
	reprint,
	nofootinbib,
	amsmath,
	amssymb,
	aps,
]{revtex4-2}

\usepackage{showyourwork}
\usepackage{orcidlink}

\usepackage{xcolor}
\usepackage{graphicx}
\graphicspath{{figures/}}
\usepackage{dcolumn}
\usepackage{bm}
\usepackage{hyperref}
\usepackage{verbatim}
\usepackage{xspace}            
\usepackage{multirow}          
\usepackage{nicematrix}
\usepackage{tikz}              
\usepackage{booktabs}
\usepackage{siunitx}
\usepackage{float}             

\usepackage{fontawesome5}

\DeclareRobustCommand{\codelink}[1]{\,\href{#1}{\mbox{\scalebox{0.65}{\faGithub}}}}

\renewcommand{\d}{{\bf d}}

\newcommand{\btheta}{\mbox{\boldmath $\theta$}}
\newcommand{\bdelta}{\mbox{\boldmath $\delta$}}

\renewcommand{\S}{{\bf S}}
\newcommand{\A}{{\bf A}}
\newcommand{\Y}{{\bf Y}}
\newcommand{\Z}{{\bf Z}}

\newcommand{\AET}{\textsc{aet}\xspace}
\newcommand{\XYZ}{\textsc{xyz}\xspace}

\newcommand{\tr}{\mbox{tr}}
\newcommand{\I}{\mbox{{\bf I}}}
\newcommand{\e}{{\rm e}}
\newcommand{\M}{{\bf M}}
\newcommand{\U}{{\bf U}}
\renewcommand{\u}{\mathbf {u}}
\renewcommand{\v}{{\bf v}}

\hypersetup{
    colorlinks=true,
    linkcolor=blue!70!black,  
    citecolor=green!50!black, 
    urlcolor=blue!50!white    
}

\AtBeginDocument{\renewcommand{\selectlanguage}[1]{}}

\begin{document}
	\preprint{APS/123-QED}

	\title{Multivariate Bayesian P-spline estimation of spectral density matrices,\\
		with application to LISA TDI noise}

    \author{Avi Vajpeyi$^{1}$\orcidlink{0000-0002-4146-1132}}
    \author{Renate Meyer$^{1}$\orcidlink{0000-0003-0268-8569}}
    \author{Patricio Maturana-Russel$^{1,2}$\orcidlink{0000-0002-5211-9818}}
    \author{Jianan Liu$^{1}$\orcidlink{0009-0007-4718-1215}}
    
    \affiliation{$^{1}$Department of Statistics, The University of Auckland, Auckland, New Zealand}
    \affiliation{$^{2}$Department of Mathematical Sciences, Auckland University of Technology, Auckland, New Zealand}

	\date{\today}

	\begin{abstract}
		We present a Bayesian P-spline method for estimating the
    frequency-dependent cross-spectral density matrix of stationary
    multivariate time series. The inverse spectral matrix is parametrised
    through its frequency-varying Cholesky decomposition, which
    guarantees Hermitian positive definiteness at every frequency. Each
    real log-diagonal entry and each real and imaginary off-diagonal
    entry is given an independent penalised B-spline prior that controls
    smoothness. Inference uses a blocked, coarse-grained Whittle
    likelihood with safe-Bayes $\eta$-tempering to stabilise posterior
    calibration, sampled by the No-U-Turn Sampler from a variational
    initialisation. On synthetic VAR(2) benchmarks with known ground
    truth, the method recovers both diagonal and cross-spectral
    structure, attains near-nominal credible-interval coverage, and
    achieves a relative integrated squared (Frobenius) error (RISE) that
    decreases with sample size. We then apply the method to publicly
    released simulated LISA time-delay interferometry (TDI) data in two
    noise configurations. In the idealised symmetric case, the full
    multivariate model and a reduced model that assumes a diagonal
    \AET\ noise covariance agree to within $\sim10^{-3}$ in RISE.
    Under realistic noise that is asymmetric across the six Movable
    Optical Sub-Assemblies (MOSAs), the \AET-diagonal
    assumption fails by more than an order of magnitude in RISE
    ($\sim\!3.3\!\times\!10^{-2}$ versus $\sim\!10^{-3}$), whereas the
    full multivariate model recovers the cross-spectral structure.
	\end{abstract}


	\maketitle

	\section{\label{sec:intro}Introduction}

	Accurate estimation of the noise power spectral density (PSD) underpins
    gravitational-wave (GW) data analysis: search, parameter estimation, and
    stochastic-background inference all require a reliable noise model across
    the detector band. For LIGO and Virgo, parametric models augmented with
    spline components, most notably BayesLine~\citep{littenberg_bayesian_2015}, are widely used.
    The space-based detector LISA~\citep{Amaro-SeoanePau2017LISA} poses a harder problem: its three
    time-delay interferometry (TDI) channels are correlated, and their noise
    properties vary smoothly but non-trivially across the millihertz band.
    A single-channel noise model cannot capture the cross-spectral correlations
    that matter for joint inference.
     
    Flexible nonparametric Bayesian approaches to PSD estimation have a well-developed
    statistical foundation, characterised by an expansion of the PSD in terms of a
    potentially large or infinite number of basis functions combined with noninformative priors on the
    coefficients. Examples include
    Bernstein polynomials with weights induced by Dirichlet process priors~\citep{beyond_whittle},
    cosines with a discounted regularized horseshoe prior~\citep{jianan_vi_sgvb_et}, and
    B-splines with knots and weights induced by Dirichlet process priors~\citep{edwards_bayesian_2019}.
    With an a priori unknown and potentially infinite number of basis functions, posterior
    computation requires trans-dimensional Markov chain Monte Carlo (MCMC) algorithms,
    which are computationally expensive. Penalised B-splines (P-splines)~\citep{EilersMarx1996}
    avoid the trans-dimensional step by representing smooth functions as linear combinations
    of a finite but large number of B-spline basis functions, with a roughness penalty on
    adjacent coefficients that prevents overfitting. In a Bayesian formulation this
    penalty can be incorporated naturally into the prior distribution of the
    spline coefficients. Applied to
    spectral density estimation via the Whittle likelihood, this yields a
    smooth, data-driven estimator that requires no parametric noise template.
    Univariate Bayesian P-spline PSD methods were introduced by
    \citet{maturana-russel_bayesian_2021}. They have since been applied to GW
    noise estimation, including recent LISA-specific work that uses log-spline
    representations tuned to the steep low-frequency rise of LISA noise
    \citep{aimen2026lisa}. Related nonparametric strategies model unknown noise
    shapes in stochastic-background searches with other spline families rather
    than penalised B-splines. These include cubic B-spline single-link noise
    models analysed by Bayesian model selection \citep{baghi_uncovering_2023},
    and Akima-spline noise components introduced at the TDI level with the
    number and location of knots inferred by a reversible-jump MCMC
    \citep{santini2025sgwb}.
     
    These univariate approaches model each channel independently. This is
    justified in the idealised equal-arm stationary LISA configuration, which
    has vanishing off-diagonal cross-spectra between the three TDI channels,
    but realistic noise --- with unequal arm lengths, noise-level imbalance
    across the six Movable Optical Sub-Assemblies (MOSAs), and
    instrumental breathing~\citep{muratore2021instrumental} --- generates
    non-trivial off-diagonal entries in the $3\times 3$ spectral density
    matrix $\S(f)$, whose normalised magnitudes (the channel-pair
    coherences) quantify the cross-channel correlation that
    single-channel models discard. Treating these cross-spectra as zero when they are not
    can degrade sensitivity to stochastic GW backgrounds and bias parameter
    estimation~\citep{CiredduLnLforNetwork,sgwb-unequal-noise-lisa-olaf}.
    Any extension of univariate methods to nonparametrically model $\S(f)$
    needs to ensure Hermitian positive definiteness
    at every frequency while allowing each matrix element to vary smoothly
    with frequency.
     
    Existing nonparametric treatments of LISA TDI spectral matrices follow two
    broad routes. \citet{muratore_impact_2023} model the diagonal PSDs and the
    real and imaginary parts of the cross-spectra of $\S(f)$ as smooth
    fractional deviations from a design spectrum, each represented by a natural
    cubic spline. They note that this construction is not guaranteed to yield a
    positive-definite matrix away from the design point, and that it is well
    suited to their local Fisher-matrix study of noise-knowledge uncertainty
    rather than to fitting data directly. A second route models the single-link
    noise log-PSD nonparametrically with cubic B-splines and maps it to the TDI
    variables through the known TDI transfer functions~\citep{baghi_uncovering_2023}.
    Because the resulting TDI covariance is built from a valid single-link
    covariance through this linear transfer, it is Hermitian positive definite
    by construction, at the cost of assuming the simplified single-link noise
    structure.
     
    In this paper, we model the PSD matrix of the \XYZ TDI variables directly by extending the Bayesian P-spline PSD framework
    \citep{maturana-russel_bayesian_2021} to the multivariate $p$-dimensional
    setting. We parametrise $\S(f)^{-1}$ through its Cholesky
    decomposition~\citep{RosenOri2007Aeom, Hu2023}, which factorises the
    posterior density into $p$ independent components that can be
    sampled in parallel and guarantees the Hermitian positive definiteness
    condition.
    Each block's log-diagonal spectra and real and imaginary parts of the complex
    off-diagonal cross-spectra are modelled with penalised B-splines under a
    hierarchical smoothing prior.
    This prior is updated by a blocked, coarse-grained Whittle likelihood
    with safe-Bayes $\eta$-tempering~\citep{grunwald2019} to stabilise
    posterior calibration. Each chain is initialised from a draw of a
    stochastic variational approximation~\citep{hoffman2013_SVI}, with
    sampling performed by the No-U-Turn Sampler~\citep{hoffman2014_NUTS}.
    We validate the method on a three-channel VAR(2) simulation with known
    ground truth, and on publicly released simulated LISA TDI datasets in a
    symmetric and a physically realistic asymmetric per-MOSA noise configuration. In the
    asymmetric case an \AET-restricted model fails by more than an order of magnitude in
    relative integrated squared (Frobenius) error while the full multivariate model
    recovers the cross-spectral structure. A further benchmark against the
    bivariate estimators of \citet{yixuan_vnpc} on a VAR(2) problem with
    closed-form ground truth is reported in Appendix~\ref{app:var2_appendix}.
     
    The remainder of the paper is organised as follows.
    Section~\ref{sec:psplines} develops the statistical model and the
    inference scheme. Section~\ref{sec:application} presents simulation
    studies on a 3D VAR(2) benchmark and on publicly released simulated LISA TDI data.
    Section~\ref{sec:discussion} summarises the results, discusses limitations
    and future directions.

	\section{Multivariate P-splines}\label{sec:psplines}

	\subsection{Likelihood}
	A glossary of the symbols used throughout this paper is provided in
	Table~\ref{tab:definitions} (Appendix~\ref{app:notation}). Readers
	may find it useful to keep this table open while reading the rest of
	this section. Let
	$\Z=(\Z_{1},\ldots,\Z_{n})^{\top}\in \mathbb{R}^{n\times p}$ be a
	$p$-dimensional stationary, mean-zero time series sampled at intervals
	$\Delta_{t}=1/(2f_{Ny})$, so that $\Z_{t}=\Z(t\Delta_{t})$ for
	$t=1,\ldots,n$, where $f_{Ny}$ is the Nyquist frequency. The total
	observation time is $T=n\Delta_{t}$ and the frequency resolution is
	\begin{align}
		\Delta_{f} = \frac{1}{n \Delta_{t}}= \frac{1}{T}\, .
	\end{align}
	The discrete Fourier transform (DFT) of $\Z$ is
	\begin{align}
		\d(f_{k}) = \Delta_{t}\sum_{t=1}^{n}\Z_{t}\exp \left(-2\pi i \frac{k}{n}t \right)\, ,
	\end{align}
	with $f_{k}=k/T$ for $k=1,\ldots,N$, where $N=\lfloor n/2\rfloor$.

	For stationary series with absolutely summable matrix autocovariances
	$\sum_{\tau=-\infty}^{\infty}\lVert\Gamma(\tau)\rVert<\infty$, the Fourier coefficients
	$\d(f_{k})$ are asymptotically independent, complex Gaussian with mean zero
	and covariance $T\,\S(f_{k})$, where
	\[
		\S(f)=\frac{1}{2f_{Ny}}\sum_{\tau=-\infty}^{\infty}\Gamma(\tau) \exp
		\left(-2\pi i f \tau\Delta_{t} \right)
	\]
	is the two-sided spectral density matrix --- the Fourier transform of the
	autocovariance $\Gamma(\tau)=\mathbb{E}(\Z_{t}\Z_{t+\tau}^{\top})$ at
	integer lag $\tau$. While the likelihood below is expressed in terms of
	the two-sided density, all spectra reported in this paper (estimates,
	analytic references, and figures) follow the one-sided convention,
	obtained by doubling the two-sided density at positive frequencies. This
	asymptotic Gaussian approximation is the basis of the multivariate
	Whittle likelihood~\citep{grainger2023multivariate}
	\begin{align}
		\label{eq:Whittle likelihood}
		\mathcal{L}(\d|\S) & \propto \prod_{k=1}^{N}\left|\S(f_{k})\right|^{-1}\times \nonumber         \\
		                                                & \exp\left(-\frac{1}{T}\d(f_{k})^{*} \S(f_{k})^{-1}\d(f_{k})\right),
	\end{align}
	where $\d(f_{k})^{*}$ denotes the conjugate transpose of $\d(f_{k})$
	and $\left|\cdot\right|$ denotes the matrix determinant. In general a
	spectral density matrix is only guaranteed to be Hermitian positive
	semidefinite. The Whittle likelihood, however, requires
	$\S(f_{k})^{-1}$ to exist, so we assume strict positive definiteness
	throughout (a property the Cholesky parametrisation of
	Section~\ref{subsec:cholesky} enforces by construction).

	The diagonal entries $S_{ii}(f)$ are the power spectral densities (PSDs) of channel $i$,
	while the off-diagonals $S_{ij}(f)$ for $i\neq j$ are the (complex)
	cross-spectra. A scale-free summary of the linear dependence between
	channels $i$ and $j$ at frequency $f$ is the complex coherence,
	\begin{align}
		\label{eq:coherence}
		C_{ij}(f) \;=\; \frac{S_{ij}(f)}{\sqrt{S_{ii}(f)\,S_{jj}(f)}},
		\qquad |C_{ij}(f)|\in[0,1],
		\codelink{https://github.com/nz-gravity/LogPSplinePSD/blob/v0.1.0/src/log_psplines/datatypes/multivar_utils.py\#L213}
	\end{align}
	whose magnitude is zero when channels $i$ and $j$ are uncorrelated at
	frequency $f$ and unity when they are perfectly linearly related.
	We use $|C_{ij}(f)|$ to visualise off-diagonal recovery in
	the LISA application (Sec.~\ref{subsec:lisa}).

\begin{figure*}[!t]
		\centering
		\includegraphics[width=\linewidth]{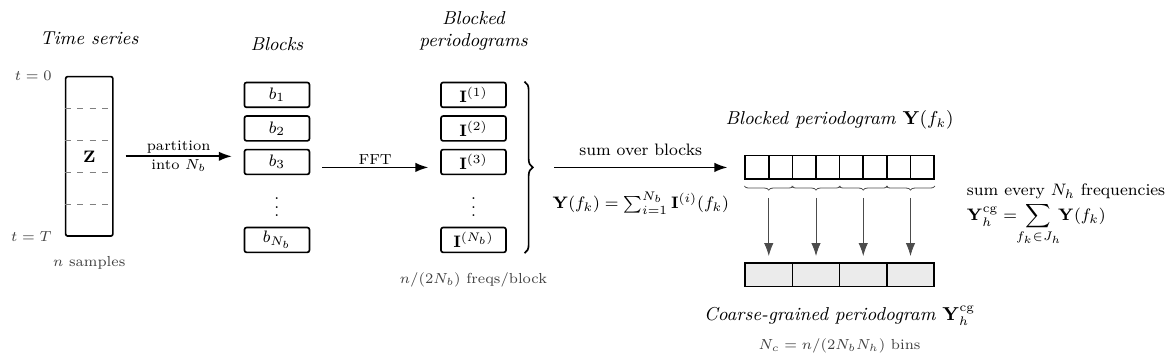}
		\caption{Construction of the blocked, coarse-grained periodogram.
			The $n$-sample time series is partitioned into $N_{b}$
			non-overlapping blocks $\{b_{i}\}$ of length $T/N_{b}$. Each
			block is Fourier-transformed independently, yielding $N_{b}$
			single-block periodograms $\mathbf{I}^{(i)}$ each of length
			$n/(2N_{b})$. These are summed across blocks to form the
			blocked periodogram
			$\mathbf{Y}(f_{k})=\sum_{i=1}^{N_{b}}\mathbf{I}^{(i)}(f_{k})$
			(related to the Welch estimator by
			$\bar{\mathbf{I}}=\mathbf{Y}/N_{b}$), which is then summed
			over consecutive groups of $N_{h}$ adjacent fine frequencies
			to produce the final coarse-grained periodogram
			$\mathbf{Y}_{h}^{\mathrm{cg}}$ of length $n/(2N_{b}N_{h})$.}
		\label{fig:blocked_likelihood}
	\end{figure*}

	\paragraph*{Blocking.}
	To trade frequency resolution for periodogram stability we partition
	$\Z$ into $N_{b}$ equal, non-overlapping blocks
	$(\Z^{(1)},\ldots,\Z^{(N_{b})})$, each of duration $T_{b}=T/N_{b}$
	and containing $n/N_{b}$ samples (illustrated in
	Fig.~\ref{fig:blocked_likelihood}). Each block may be tapered by a
	window function $w_{t}$ before the DFT to suppress spectral leakage.
	Writing the number of samples per block as $n_{b}=n/N_{b}$, we define
	the window's equivalent noise bandwidth as
	\begin{equation}
		\label{eq:enbw}
		\mathrm{ENBW}
		\;=\;
		n_{b}\frac{\sum_{t=1}^{n_b}w_{t}^{2}}
		{\left(\sum_{t=1}^{n_b}w_{t}\right)^{2}}\, .
		\codelink{https://github.com/nz-gravity/LogPSplinePSD/blob/v0.1.0/src/log_psplines/datatypes/multivar.py\#L248}
	\end{equation}
	We rescale each block periodogram by
	$1/\mathrm{ENBW}$, so that the complex-Wishart mean of $\Y(f_{k})$
	remains $T_{b}\,\S(f_{k})$. The separate effect of tapering on the
	effective likelihood information is described below.
	Denoting the DFT of block $i$ by
	$\d^{(i)}$, and assuming blocks are approximately independent under
	stationarity, the blocked Whittle likelihood factorises as
	\begin{equation}
		\label{eq:block_lnl}
		\mathcal{L}_{b}(\d|\S) = \prod_{i=1}^{N_b}\mathcal{L}(\d^{(i)}|\S)\, .
	\end{equation}
	Collecting the sum over blocks into a single Hermitian matrix
	\begin{equation}
		\label{eq:Y_def}
		\Y(f_{k}) \;=\; \sum_{i=1}^{N_{b}}\I^{(i)}(f_{k})
		\;=\; N_{b}\,\bar{\I}(f_{k}),
	\end{equation}
	where $\I^{(i)}(f_{k})=\d^{(i)}(f_{k})\d^{(i)}(f_{k})^{*}$ is the
	block-$i$ periodogram and
	$\bar{\I}(f)=N_{b}^{-1}\sum_{i=1}^{N_{b}}\I^{(i)}(f)$ is the
	block-averaged periodogram (the Welch estimator), and applying the
	trace cyclic property $\tr(ABC)=\tr(CAB)$, the blocked likelihood
	reduces to
	\begin{align}
		\label{eq:blocked-trace}
		\mathcal{L}_{b}(\d|\S) & \propto \prod_{k=1}^{n/(2N_{b})}
		\left|\S(f_{k})\right|^{-N_{b}}\times \nonumber \\
		                       & \exp\!\left(-\tfrac{1}{T_{b}}\tr\!\left[\S(f_{k})^{-1}\Y(f_{k})\right]\right).
	\end{align}

	Under the asymptotic Gaussian approximation, $\Y(f_{k})$ follows a
	complex Wishart distribution\footnote{We write
	$\A\sim\mathcal{CW}_{d}(\M,m)$ if the $d\times d$ Hermitian positive
	definite matrix $\A$ has density proportional to
	$\left|\M\right|^{-m}\exp(-\tr[\M^{-1}\A])$, with mean $m\M$.}
	from summing $N_{b}$ assumed-independent single-block periodogram
	contributions:
	\begin{align}
	\I^{(i)}(f_{k})&\sim\mathcal{CW}_{p}(T_{b}\S(f_{k}),1),\\
	\Y(f_{k})&\sim\mathcal{CW}_{p}(T_{b}\S(f_{k}),N_{b}).
	\end{align}
	\paragraph*{Eigendecomposition.}
	Because $\Y(f_{k})$ is Hermitian positive semidefinite, it admits an
	eigendecomposition
	$\Y(f_{k})=\sum_{\nu=1}^{p}\lambda^{(k)}_{\nu}\v^{(k)}_{\nu}\v^{(k)*}_{\nu}
	=\sum_{\nu=1}^{p}\u^{(k)}_{\nu}\u^{(k)*}_{\nu}$,
	with rescaled eigenvectors
	$\u^{(k)}_{\nu}=\sqrt{\lambda^{(k)}_{\nu}}\,\v^{(k)}_{\nu}$.
	Substituting into the trace in \eqref{eq:blocked-trace} gives the
	eigenvector form of the blocked likelihood,
	\begin{align}
		\mathcal{L}_{b}(\d|\S)
		&\propto \prod_{k=1}^{n/(2N_{b})}
		\left|\S(f_{k})\right|^{-N_{b}} \nonumber\\
		&\quad\times
		\exp\!\left(-\tfrac{1}{T_{b}}\sum_{\nu=1}^{p}
		\u_{\nu}^{(k)*}\S(f_{k})^{-1}\u_{\nu}^{(k)}\right).
		\label{eq:log-like}
	\end{align}
	This follows from the spectral decomposition of the Hermitian positive
	semidefinite $\Y(f_k)$ and the linearity of the trace. The
	eigenvector representation is used computationally because it turns the
	trace contribution into a sum of quadratic forms in the Cholesky
	parameters, which is the form that factorises into the per-channel
	regressions below.

	\paragraph*{Coarse graining.}
	Blocking reduces the number of periodogram matrices from $N$ to
	$n/(2N_{b})$, but each of the remaining frequencies still requires a
	separate evaluation of $\S(f_{k})^{-1}$. Where $\S(f)$ is slowly
	varying, we can pool adjacent frequencies to gain further speed
	(illustrated in Fig.~\ref{fig:blocked_likelihood}). Partition the block frequencies into
	$N_{c}$ disjoint subsets
	$\{f_{1},\ldots,f_{n/(2N_{b})}\}=\bigcup_{h=1}^{N_{c}}J_{h}$, each
	containing $N_{h}$ adjacent frequencies with bin centre
	$\bar{f}_{h}\equiv \tfrac{1}{N_{h}}\sum_{f\in J_{h}} f$, which need
	not coincide with a Fourier frequency.
	Replace the per-frequency matrix $\Y(f_{k})$ by the within-bin sum
	$\Y_{h}^{\mathrm{cg}}=\sum_{f\in J_{h}}\Y(f)$. 

	Under the assumption that $\S(f)\approx\S(\bar{f}_{h})$ across each $J_{h}$,
	the $\Y(f)$ within the bin are approximately i.i.d.\
	$\mathcal{CW}_{p}(T_b\S(\bar{f}_{h}), N_{b})$, and by the additivity of the
	complex Wishart under a common scale matrix,
	\begin{equation}
		\Y_{h}^{\mathrm{cg}}\;\dot\sim\;\mathcal{CW}_{p}\!\left(T_b\,\S(\bar{f}_{h}),\, N_{b}N_{h}\right).
		\codelink{https://github.com/nz-gravity/LogPSplinePSD/blob/v0.1.0/src/log_psplines/preprocessing/coarse_grain.py\#L177}
	\end{equation}
	The approximation is only as good as the constant-$\S$ assumption, so bin
	edges must avoid sharp spectral features (peaks, lines, transfer-function
	nulls).

	The coarse-grained likelihood then takes the same form as
	Eq.~\eqref{eq:log-like} with the $n/(2N_{b})$ block frequencies
	replaced by $N_{c}$ coarse bins and a Wishart shape parameter
	$N_{b}N_{h}$ (the number of fine periodograms summed per coarse
	bin):
	\begin{align}
		\label{eq:log-like-coarse}
		\mathcal{L}_{b,c}(\d|\S) & \propto \prod_{h=1}^{N_{c}}
		\left|\S(\bar{f}_{h})\right|^{-N_{b}N_{h}}\times \nonumber \\
		                         & \exp\!\left(-\tfrac{1}{T_{b}}\tr\!\left[\S(\bar{f}_{h})^{-1}\Y_{h}^{\mathrm{cg}}\right]\right),
		                         \codelink{https://github.com/nz-gravity/LogPSplinePSD/blob/v0.1.0/src/log_psplines/pipeline/models.py\#L197}
	\end{align}
	with eigendecomposition
	\begin{align}
	\Y_{h}^{\mathrm{cg}}&=\sum_{\nu=1}^{p} \u^{(h)}_{\nu}\u^{(h)*}_{\nu}\, ,\\
	\u^{(h)}_{\nu}&=\sqrt{\lambda^{(h)}_{\nu}}\,\v^{(h)}_{\nu}\, .
	\end{align}

	Although rescaling the periodogram by $1/\mathrm{ENBW}$ preserves its mean,
	tapering correlates neighbouring Fourier coefficients and therefore reduces
	the effective number of independent frequency-domain observations. We
	approximate this loss of information by dividing the log-likelihood by the
	window's equivalent noise bandwidth in Eq.~\eqref{eq:enbw}, or equivalently
	by using the power likelihood
	\begin{equation}
		\label{eq:enbw-likelihood}
		\mathcal{L}_{b,c,w}(\d\mid\S)
		\;\propto\; \mathcal{L}_{b,c}(\d\mid\S)^{1/\mathrm{ENBW}}\, .
		\codelink{https://github.com/nz-gravity/LogPSplinePSD/blob/v0.1.0/src/log_psplines/pipeline/models.py\#L241}
	\end{equation}
	Thus a rectangular window leaves the likelihood unchanged, whereas a
	non-rectangular taper reduces its curvature by the factor
	$1/\mathrm{ENBW}$.

	\subsection{Parametrisation using the Cholesky Decomposition}
	\label{subsec:cholesky}
	Following~\citet{RosenOri2007Aeom,Hu2023}, we parametrise the
	\emph{inverse} spectral density matrix through its Cholesky decomposition.
	Working with $\S^{-1}$ rather than $\S$ is what makes the problem
	tractable: the quadratic form $\u^{*}\S^{-1}\u$ splits into a sum of $p$
	decoupled quadratic terms, which in turn factorises the likelihood into
	$p$ independent regressions (one per channel) that can be sampled in
	parallel. We write
	\begin{equation}
		\mathbf{S}(f_{k})^{-1}= \mathbf{T}_{k}^{*} \, \mathbf{D}_{k}^{-1}\, \mathbf{T}_{k},
	\end{equation}
	where $\mathbf{D}_{k}$ is diagonal with positive entries
	$\delta_{1k}^{2}, \delta_{2k}^{2}, \ldots, \delta_{pk}^{2}$, and
	\begin{align}
		\mathbf{T}_{k} = \begin{pmatrix}1&0&0&\cdots&0 \\ -\theta_{21}^{(k)}&1&0&\cdots&0 \\ -\theta_{31}^{(k)}&-\theta_{32}^{(k)}&1&\ddots&\vdots \\ \vdots&\vdots&\ddots&\ddots&0 \\ -\theta_{p1}^{(k)}&-\theta_{p2}^{(k)}&\cdots&-\theta_{p,p-1}^{(k)}&1\end{pmatrix}
	\end{align}
	is a $p \times p$ complex unit lower triangular matrix with
	$\theta_{il}^{(k)}$ representing the complex-valued off-diagonal elements for
	$i > l$.

	This parametrisation factorises the likelihood into a product of $p$
	per-channel terms,
	\begin{equation}
		\mathcal{L}(\d|\S) \propto \prod_{j=1}^{p}\mathcal{L}_{j}(\btheta_{j},\bdelta_{j}),
		\label{eq:likelihood}
	\end{equation}
	one for each row of $\mathbf{T}$. Each $\mathcal{L}_{j}$ is a
	univariate Whittle regression of the $j$th channel of the eigenvector
	pseudo-observations onto the preceding $(j-1)$ channels:
	\begin{widetext}
	\begin{equation}
		\label{eq:likelihoodj}
		\mathcal{L}_{j}(\btheta_{j},\bdelta_{j}) \propto
		\prod_{h=1}^{N_{c}} \delta_{jh}^{-2N_{b}N_{h}}
		\exp\!\left(\frac{-\sum_{\nu=1}^{p}\left|u_{j\nu}^{(h)}-\sum_{l=1}^{j-1}\theta_{jl}^{(h)}\,u_{l\nu}^{(h)}\right|^{2}}
		              {T_{b}\,\delta_{jh}^{2}}\right), \codelink{https://github.com/nz-gravity/LogPSplinePSD/blob/v0.1.0/src/log_psplines/pipeline/models.py\#L81}
	\end{equation}
	\end{widetext}%
	with regression coefficients $\theta_{jl}^{(h)}$ (the off-diagonal
	Cholesky entries at coarse bin $h$) and noise variance
	$\delta_{jh}^{2}$ (the $j$th diagonal entry of $\mathbf{D}_{h}$).
	The eigenvectors $\u_{\nu}^{(h)}$ are those of the coarse-grained
	statistic $\Y_{h}^{\mathrm{cg}}$, evaluated at the bin midpoint $\bar f_{h}$.
	The parameter blocks $\btheta_{j}$ and $\bdelta_{j}$ collect all
	$\theta_{jl}^{(h)}$ ($l<j$) and $\delta_{jh}$ across the coarse bins
	$h=1,\ldots,N_{c}$.

	In particular, for $p=3$ channels (as in the LISA case), the three
	factors depend on the parameter subvectors $\bdelta_{1}$,
	$(\bdelta_{2},\btheta_{2})$, and $(\bdelta_{3},\btheta_{3})$,
	respectively, as follows:
	\begin{widetext}
	\begin{align}
		\mathcal{L}_{1}(\bdelta_{1})
		&\propto \prod_{h=1}^{N_{c}}\delta_{1h}^{-2N_{b}N_{h}}
		  \exp\!\left(\frac{-\sum_{\nu=1}^{p}\left|u_{1\nu}^{(h)}\right|^{2}}
		              {T_{b}\,\delta_{1h}^{2}}\right), \label{eq:likelihood1}\\
		\mathcal{L}_{2}(\btheta_{2},\bdelta_{2})
		&\propto \prod_{h=1}^{N_{c}}\delta_{2h}^{-2N_{b}N_{h}}
		  \exp\!\left(\frac{-\sum_{\nu=1}^{p}\left|u_{2\nu}^{(h)}-\theta_{21}^{(h)}u_{1\nu}^{(h)}\right|^{2}}
		              {T_{b}\,\delta_{2h}^{2}}\right), \label{eq:likelihood2}\\
		\mathcal{L}_{3}(\btheta_{3},\bdelta_{3})
		&\propto \prod_{h=1}^{N_{c}}\delta_{3h}^{-2N_{b}N_{h}}
		  \exp\!\left(\frac{-\sum_{\nu=1}^{p}\left|u_{3\nu}^{(h)}-\theta_{31}^{(h)}u_{1\nu}^{(h)}-\theta_{32}^{(h)}u_{2\nu}^{(h)}\right|^{2}}
		              {T_{b}\,\delta_{3h}^{2}}\right). \label{eq:likelihood3}
	\end{align}
	\end{widetext}%
	Note that a factorised likelihood does not in general imply a factorised
	posterior. Here, however, the posterior also factorises because we place independent
	priors across these three parameter blocks,
	\begin{align}
		\pi(\bdelta_{1},\bdelta_{2},\btheta_{2}, & \bdelta_{3},\btheta_{3}) = \notag                                                                         \\
		                                         & \pi_{1}(\bdelta_{1})\,\pi_{2}(\bdelta_{2},\btheta_{2})\,\pi_{3}(\bdelta_{3},\btheta_{3}),                             \\
		p(\bdelta_{1},\bdelta_{2},\btheta_{2},   & \bdelta_{3},\btheta_{3}\mid \d) \propto \notag                                    \\
		                                         & \mathcal{L}_{1}(\bdelta_{1})\,\pi_{1}(\bdelta_{1})\times \notag                                  \\
		                                         & \mathcal{L}_{2}(\bdelta_{2},\btheta_{2})\,\pi_{2}(\bdelta_{2},\btheta_{2})\times \notag \\
		                                         & \mathcal{L}_{3}(\bdelta_{3},\btheta_{3})\,\pi_{3}(\bdelta_{3},\btheta_{3}).
	\end{align}
	Hence Bayes' rule yields three independent posterior factors corresponding to $\bdelta
	_{1}$, $(\bdelta_{2},\btheta_{2})$, and $(\bdelta_{3},\btheta_{3})$. This
	means we can run an MCMC chain for each block separately and in parallel, which
	in general reduces the computation time.

	\subsection{Safe-Bayes $\eta$-tempering}
	\label{subsec:eta}
	The Whittle likelihood is an asymptotic approximation to the exact Gaussian
	likelihood of $\Z$, and the coarse-grained form in
	Eq.~\eqref{eq:log-like-coarse} compounds this with a within-bin
	constant-$\S$ assumption. Both approximations introduce mild
	misspecification. A well-known consequence of likelihood misspecification is
	that the posterior can over-concentrate, producing credible intervals that
	under-cover the truth even when the posterior mode is accurate
	\citep{grunwald2019}.

	We mitigate this with the safe-Bayes correction of
	\citet{grunwald2019}: raise the likelihood to a power $\eta\in(0,1]$,
	\begin{equation}
		\label{eq:eta-tempered}
		\mathcal{L}_{\eta}(\d\mid\S)
		\;\propto\; \mathcal{L}_{b,c,w}(\d\mid\S)^{\eta}
		\;\propto\; \mathcal{L}_{b,c}(\d\mid\S)^{\eta/\mathrm{ENBW}},
		\codelink{https://github.com/nz-gravity/LogPSplinePSD/blob/v0.1.0/src/log_psplines/pipeline/models.py\#L244}
	\end{equation}
	which, together with the ENBW correction, gives an effective scaling
	$N_{b}N_{h}\mapsto \eta N_{b}N_{h}/\mathrm{ENBW}$ in
	Eq.~\eqref{eq:log-like-coarse}. The limit $\eta\to 1$ recovers the
	window-corrected likelihood, while $\eta\to 0$ recovers the prior. Intermediate
	$\eta$ down-weights the (asymptotic) likelihood relative to the prior in
	exactly the right amount to restore nominal coverage when the
	approximation is imperfect.

	A useful way to read this correction is through the curvature of the
	log-likelihood. From Eq.~\eqref{eq:log-like-coarse}, the observed
	Fisher information per coarse bin scales linearly with the number of
	fine periodograms summed in each bin: schematically,
	\begin{equation}
		\label{eq:fisher-scaling}
		-\,\partial^{2}_{\theta\theta}\,\log\mathcal{L}_{b,c,w}(\d\mid\S(\theta))
		\;\propto\; \frac{N_{b}N_{h}}{\mathrm{ENBW}}\,\mathcal{I}_{0}(\theta),
	\end{equation}
	where $\mathcal{I}_{0}$ is the per-bin information matrix induced by
	a single complex Wishart observation. For a fixed window, increasing either
	blocking or coarse-graining therefore sharpens the log-posterior by the
	same factor $N_{b}N_{h}$, while tapering offsets this by
	$1/\mathrm{ENBW}$. We treat $\eta$ as a tunable calibration
	knob rather than a fundamental quantity. For the small synthetic
	benchmarks in Section~\ref{subsec:sim_3d_var2}
	($n=16{,}384$, $N_{b}N_{h}\!\lesssim\!10^{2}$) the un-tempered limit
	$\eta=1$ is already well-calibrated and is used throughout. The
	correction becomes important for the LISA application
	(Section~\ref{subsec:lisa}), where $N_{b}N_{h}$ is two to three orders
	of magnitude larger and the resulting log-likelihood curvature drives
	both posterior over-concentration and, at sufficiently large
	$N_{b}N_{h}$, sampler pathologies (maximum tree-depth saturation in
	NUTS). An $\eta$-sensitivity study reported in
	Appendix~\ref{app:eta_sweep} motivates the choice of $\eta$ used
	for the LISA results.

	\subsection{P-spline modelling}
	We model the frequency-dependent Cholesky parameters with penalised
	B-splines,
	\begin{align}
		\log \delta_{jk}^{2}    & = \sum_{m=1}^{K_j}B_{m}(f_{k}) w_{j,m}^{(\delta)}\label{eq:delta_spline}     \\
		\Re[\theta_{jl}(f_{k})] & = \sum_{m=1}^{K_{jl}}B_{m}(f_{k}) w_{jl,m}^{(\Re)}\label{eq:theta_re_spline} \\
		\Im[\theta_{jl}(f_{k})] & = \sum_{m=1}^{K_{jl}}B_{m}(f_{k}) w_{jl,m}^{(\Im)}\label{eq:theta_im_spline} \codelink{https://github.com/nz-gravity/LogPSplinePSD/blob/v0.1.0/src/log_psplines/psplines/psplines.py\#L545}
	\end{align}
	where $B_{m}(f_{k})$ are B-spline basis functions evaluated at frequency
	$f_{k}$, $K_{j}$ and $K_{jl}$ are the numbers of basis functions for the
	$j$-th diagonal and $(j,l)$ off-diagonal components, respectively, and
	$w_{j,m}^{(\delta)}$, $w_{jl,m}^{(\Re)}$, $w_{jl,m}^{(\Im)}$ are the
	corresponding spline coefficients. When the blocked, coarse-grained
	likelihood is used, these expansions are evaluated at the coarse-bin
	centres $\bar{f}_{h}$, yielding the parameters $\delta_{jh}^{2}$ and
	$\theta_{jl}^{(h)}$ that enter Eq.~\eqref{eq:likelihoodj}.

	Knots are placed adaptively, in the spirit of
	\citet{maturana-russel_bayesian_2021}: for each Cholesky component the
	empirical periodogram is denoised, the absolute gradient of the result
	--- plus a small uniform floor so that featureless regions still
	receive knots --- is normalised to a probability mass function over
	frequency, and knots are positioned at equally spaced quantiles of its
	cumulative
	distribution.\codelink{https://github.com/nz-gravity/LogPSplinePSD/blob/v0.1.0/src/log_psplines/psplines/knots_locator/knot_locator.py\#L217}
	For the off-diagonal entries the real and imaginary parts are treated as
	separate components, with knots placed independently on
	$|\Re[\theta_{jl}(f_{k})]|$ and $|\Im[\theta_{jl}(f_{k})]|$. Knots therefore cluster where the Cholesky component
	varies rapidly and spread out in smooth regions, with no manual tuning.

	Each component $j$ has its own penalty matrix $\bold{P}_{j}$ with
	entries
	\begin{equation}
		\label{eq:Pj}
		[\bold{P}_{j}]_{lm}= \int_{0}^{1} B_{l}''(t)\, B_{m}''(t)\, \mathrm{d}t.
		\codelink{https://github.com/nz-gravity/LogPSplinePSD/blob/v0.1.0/src/log_psplines/psplines/initialisation.py\#L150}
	\end{equation}
	Unlike the finite-difference approximation
	$\bold{D}_{2}^{\top}\bold{D}_{2}$ used in the original P-spline
	formulation of~\citet{EilersMarx1996}, this integral form is exact for
	non-uniform knot sequences~\citep{Wand2008}. 
	Note that the integral
	form of $\bold{P}_{j}$ is the same for every component. 
	The $j$-subscript reflects only that the knot sequence and basis size
	$K_{j}$ are placed adaptively per component, so the resulting $K_{j}\times K_{j}$ matrix differs
	across~$j$.

	\subsection{Bayesian Inference}
	\label{sec:bayesian-inference}
	The full inference workflow proceeds in four stages. First, we place
	hierarchical smoothing priors on the spline coefficients of every
	Cholesky component. Second, we combine these priors with the
	$\eta$-tempered, blocked, coarse-grained Whittle likelihood
	(Eq.~\ref{eq:eta-tempered}) using the per-channel factorisation of
	Eq.~\eqref{eq:likelihood}. Third, we warm-start each MCMC chain from
	a draw of a stochastic variational approximation to the posterior, to
	bypass the expensive NUTS adaptation in low-mass-curvature regions.
	Fourth, we sample the warm-started posterior with NUTS and assess
	convergence with standard chain diagnostics. Each stage is
	described below.

	We place hierarchical priors on the spline coefficients to enforce smoothness.
	Each spline component (the diagonal log-variances $\log\delta_{j}^{2}$ and
	the real and imaginary parts of each off-diagonal entry $\theta_{jl}$) is
	assigned its own pair of precision hyperparameters; with a slight abuse of
	notation we index a generic such component by $j$:
	\begin{align}
		\bold{w}_{j} | \phi_{j} & \sim \mathcal{N}(\bold{0}, (\phi_{j} \widetilde{\bold{P}}_{j})^{-1})  \\
		\phi_{j} | \nu_{j}   & \sim \text{Gamma}(\alpha_{\phi}, \nu_{j} \beta_{\phi}) \\
		\nu_{j}              & \sim \text{Gamma}(\alpha_{\nu}, \beta_{\nu})
		\codelink{https://github.com/nz-gravity/LogPSplinePSD/blob/v0.1.0/src/log_psplines/pipeline/models.py\#L16}
	\end{align}
	where $\bold{w}_{j}$ is the vector of spline coefficients and
	$\phi_{j}$, $\nu_{j}$ are precision hyperparameters.\footnote{In the
	closely related univariate constructions of
	\citet{maturana-russel_bayesian_2021} and \citet{aimen2026lisa}
	this hyperparameter is denoted $\delta_{j}$. 
	We use $\nu_{j}$ here to
	avoid collision with the Cholesky diagonal entries $\delta_{jk}^{2}$
	introduced in Section~\ref{sec:psplines} above.}
	We use $\widetilde{\bold{P}}_{j}=\bold{P}_{j}+\epsilon\,\mathbf{I}$ with
	$\epsilon=10^{-6}$ to ensure the prior precision is strictly positive
	definite. Unless stated otherwise, the results below use
	$\alpha_{\phi}=\beta_{\phi}=1$ and
	$\alpha_{\nu}=\beta_{\nu}=1$.

	Combining the $\eta$-tempered likelihood
	(Eq.~\ref{eq:eta-tempered}) with these priors and the Cholesky
	factorisation of $\S^{-1}$, the posterior is sampled using the No-U-Turn
	Sampler~\citep[NUTS;][]{hoffman2014_NUTS}. The Cholesky parametrisation
	guarantees Hermitian positive definiteness of $\widehat{\S}(f)$ at every
	frequency by construction.

	\paragraph{Computational complexity.}
	The B-spline basis matrices are evaluated once at the start of a run,
	after the knot locations are fixed, and reused for every gradient
	evaluation. Each call to $\log\mathcal{L}_{\eta}$ then costs
	$\mathcal{O}(p^{2}\,N_{c}\,K)$ floating-point operations (using a common
	basis size $K = K_j = K_{jl}$ for all components):
	$N_{c}$ coarse bins, $K$ basis functions per component, and $p^{2}$ spline
	components --- the $p$ diagonal log-variances $\log\delta_{jh}^{2}$
	plus the real and imaginary parts of the $p(p-1)/2$ off-diagonal
	Cholesky entries $\theta_{jl}$. Memory is dominated by the same
	$p^{2}\,N_{c}\,K$ basis-matrix storage. In practice this
	per-evaluation cost is small compared to the fixed overheads of NUTS
	warm-up and SVI initialisation, so wall-clock runtime is dominated
	not by the gradient-evaluation cost but by the number of leapfrog
	integrator steps required to traverse the posterior. As a rough
	scale, a single evaluation of $\log\mathcal{L}_{\eta}$ on the 1-year
	LISA configuration ($p=3$, $N_{c}=1024$, $K=100$) takes a fraction
	of a millisecond on a single CPU core. Empirical end-to-end
	wall-clock numbers are reported alongside the tables in
	Section~\ref{sec:application}.

	\paragraph{Implementation.}
	We implement the model in \textsc{JAX}~\cite{jax2018github} with
	NumPyro~\cite{phan2019numphyro}, exploiting automatic differentiation
	and just-in-time (JIT) compilation for gradient-based sampling.
	Throughout we use degree-2 B-splines with second-order derivative
	penalties (the integral form of Eq.~\eqref{eq:Pj}). Positive
	precision parameters $\phi_{j}$ and $\nu_{j}$ are sampled on the
	log scale to keep the sampler away from the boundary at zero, and
	the frequency grid is standardised to $[0,1]$ for numerical
	stability of the B-spline basis evaluation. To initialise the spline
	parameters we first run stochastic variational inference
	\citep[SVI;][]{hoffman2013_SVI} with a low-rank Gaussian guide
	(rank~16), optimised with Adam~\citep{Kingma2015} at learning rate
	$10^{-3}$ for 2000
	iterations.\codelink{https://github.com/nz-gravity/LogPSplinePSD/blob/v0.1.0/src/log_psplines/pipeline/vi.py\#L134}
	Four NUTS chains are then initialised
	at independent draws from the resulting variational posterior. The
	NUTS mass matrix is adapted during 2000 warm-up steps, followed by
	an additional 2000 sampling steps per chain. These sampler settings
	are used for both the simulation study
	(Section~\ref{subsec:sim_3d_var2}) and the LISA application
	(Section~\ref{subsec:lisa}). Only the basis size $K$, the number of
	blocks $N_{b}$, and the tempering $\eta$ vary between the two.

	\paragraph{Convergence diagnostics.}
	We assess convergence by checking that the chains satisfy
	$\hat{R}<1.01$ (the rank-normalised Gelman--Rubin potential scale reduction factor~\cite{gelman1992inference, arviz_2019}), that the bulk and tail effective sample sizes both
	exceed $400$, that the energy-Bayesian fraction of missing
	information satisfies E-BFMI~$>0.3$, and that divergent transitions
	and maximum tree-depth saturation are absent or negligible. All results
	in Section~\ref{sec:application} satisfy these criteria unless
	explicitly noted otherwise.

	\section{\label{sec:application}Application}

	\subsection{Simulation study}
	\label{subsec:sim_3d_var2}

	We summarise spectral estimation accuracy with two complementary
	metrics. The first is the relative integrated squared (Frobenius)
	error (RISE),
	\begin{equation}
		\operatorname{RISE}= \frac{\int \! \left\lVert \widehat{S}(f) - S(f) \right\rVert_{F}^{2}
		\, \mathrm{d}f}{\int \! \left\lVert S(f) \right\rVert_{F}^{2} \, \mathrm{d}f}, \codelink{https://github.com/nz-gravity/LogPSplinePSD/blob/v0.1.0/src/log_psplines/diagnostics/_utils.py\#L56}
	\end{equation}
	where $\lVert \cdot \rVert_{F}$ denotes the Frobenius norm and the
	integrals are evaluated numerically using Simpson's rule over the
	periodogram frequency grid. This metric measures the bias of the
	posterior-median estimator $\widehat{S}(f)$ relative to the true
	spectrum $S(f)$. The second metric is the
	empirical coverage at level $\alpha$, defined as the fraction of
	(frequency, matrix-element) pairs for which the true values lie inside
	the pointwise $\alpha$ credible band, averaged over realisations. This
	metric measures the calibration of the credible intervals. We use
	$\alpha=0.9$
	throughout.\codelink{https://github.com/nz-gravity/LogPSplinePSD/blob/v0.1.0/src/log_psplines/diagnostics/_utils.py\#L131}

	We simulate from a three-dimensional VAR(2) process,
	\begin{align}
		\mathbf{X}_{t} = A_{1} \mathbf{X}_{t-1}+ A_{2} \mathbf{X}_{t-2}+ \boldsymbol{\varepsilon}_{t}, \qquad \boldsymbol{\varepsilon}_{t} \sim \mathcal{N}(\mathbf{0}, \Sigma),
	\end{align}
	with
	\begin{align}
		A_{1}  & = \mathrm{diag}(0.4, 0.3, 0.2),                                                   \\
		A_{2}  & = \begin{bmatrix}-0.2&0.5&0.0 \\ 0.4&-0.1&0.0 \\ 0.0&0.0&-0.1\end{bmatrix},       \\
		\Sigma & = \begin{bmatrix}0.25&0.00&0.08 \\ 0.00&0.25&0.08 \\ 0.08&0.08&0.25\end{bmatrix}.
	\end{align}
	These coefficient matrices define a stationary process: the companion
	matrix has spectral radius $0.774<1$.
    
	The one-sided theoretical spectral density matrix is
	\begin{align}
		\mathbf{S}(f) &= \frac{2}{f_{s}}\,H(f)\,\Sigma\,H(f)^{*}, \quad \text{with}\\
		H(f) &= \left(I-\sum_{\ell=1}^{2}A_{\ell}\, e^{-i2\pi f\ell/f_{s}}\right)^{-1}.
		\codelink{https://github.com/nz-gravity/LogPSplinePSD/blob/v0.1.0/src/log_psplines/example_datasets/varma_data.py\#L379}
	\end{align}

	We generate $500$ independent realisations of length $n=16{,}384$
	(with a 512-point burn-in) at sampling frequency $f_{s}=1$. Each
	realisation is fit with $K=10$ basis functions for each Cholesky
	component, while sweeping
	$N_{b}\in\{1,2,4,8,16\}$ and $N_{h}\in\{\text{off},2,4,8,16,32\}$.
	Safe-Bayes tempering (Section~\ref{subsec:eta}) is left at $\eta=1$
	throughout, with $N_{h}=\text{off}$ denoting no coarse-graining.
	A rectangular window ($\mathrm{ENBW}=1$) is applied to each block before
	computing the DFT.
	All other settings follow Section~\ref{sec:bayesian-inference}.

\begin{figure}
		\centering
		\includegraphics[width=1\linewidth]{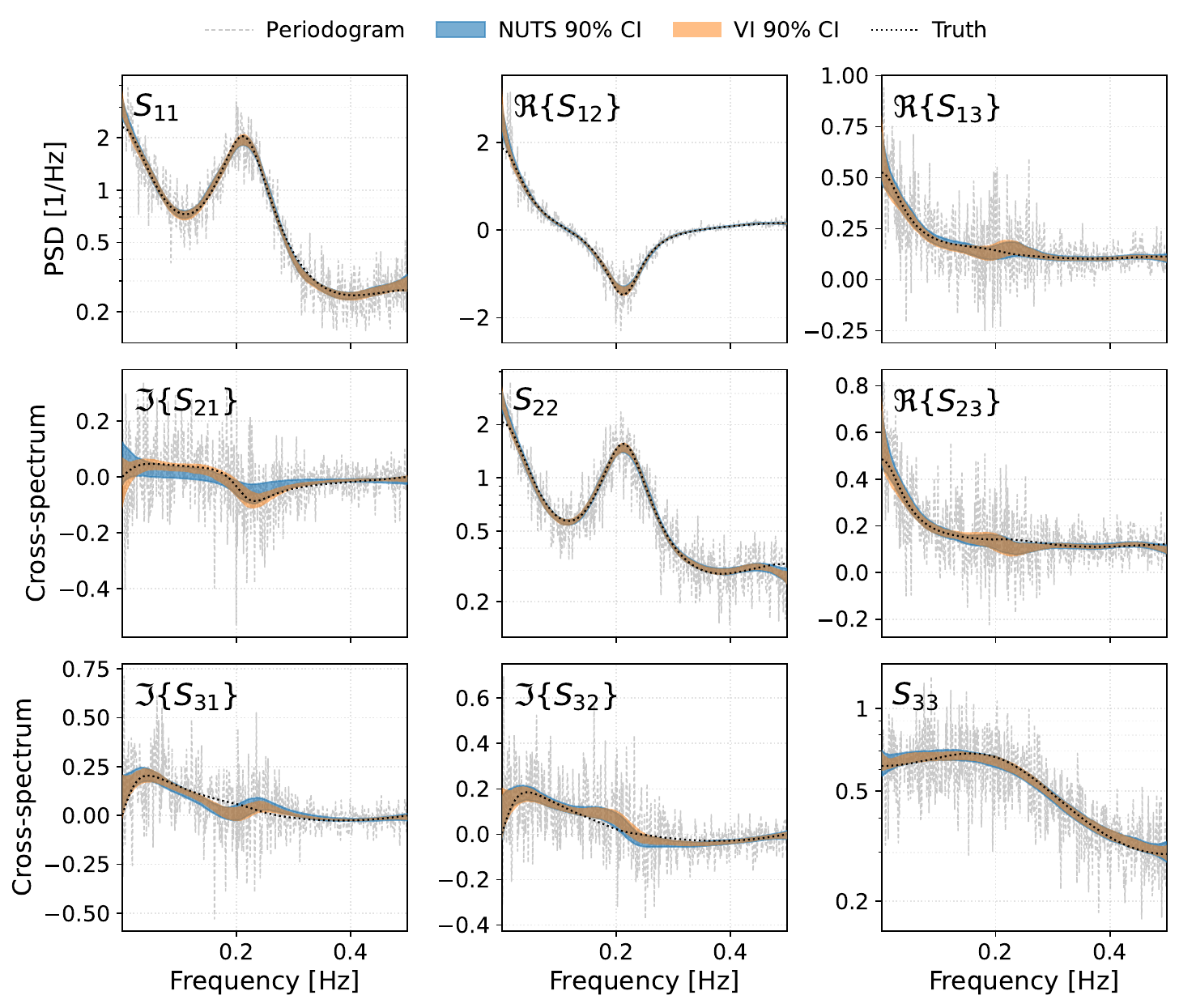}
		\caption{Representative $3\times3$ spectral density matrix estimate for a
			single $n=16{,}384$ VAR(2) realisation, fit with $K=10$, $N_{b}=4$,
			$N_{h}=4$ and $\eta=1$ (NUTS: $\mathrm{RISE}=0.114$, 90\% CI
			coverage $88\%$; SVI: $\mathrm{RISE}=0.102$, 90\% CI coverage $83\%$).
			Each panel shows the coarse-grained periodogram statistic
			$\Y_{h}^{\mathrm{cg}}/(T_{b}N_{b}N_{h})$ (grey), the theoretical spectrum
			(dotted black), and posterior medians with 90\% credible bands for
			NUTS (blue) and SVI (orange). Diagonal panels ($S_{11}$, $S_{22}$,
			$S_{33}$) show the power spectral densities, while the off-diagonal
			panels show the real and imaginary parts of the cross-spectra on
			a linear vertical scale (the VAR(2) PSDs span less than one decade,
			so all panels share a linear scale).
		}
		\label{fig:vi_vs_nuts_var3}
		\script{plot_var3.py}
	\end{figure}

	Figure~\ref{fig:vi_vs_nuts_var3} shows the full $3\times 3$ spectral
	matrix recovered for a representative $n=16{,}384$ realisation, with
	the NUTS and SVI posteriors overlaid. Both samplers recover the
	diagonal PSDs and off-diagonal cross-spectra with substantially
	smoother posterior medians than the raw periodogram, and the two
	posterior medians track each other closely ($\mathrm{RISE}=0.114$
	for NUTS versus $0.102$ for SVI). The SVI credible bands, however,
	are narrower than those of NUTS: for this realisation the
	empirical 90\% coverage drops from $88\%$ under NUTS to $83\%$ under
	SVI. This is the well-known under-dispersion of mean-field/low-rank
	Gaussian variational approximations, and it is why we use SVI only
	as a fast NUTS initialiser and retain NUTS as the production sampler
	when calibrated credible bands are required. That said, the SVI fit
	itself completes in a few seconds and recovers an accurate posterior
	median, which is attractive when only a point estimate of the
	spectrum is needed.

	Table~\ref{tab:3d_var2_results} reports a representative slice of the
	$(N_{b},N_{h})$ grid. Coverage is near-nominal (89.8--90.8\%) across
	every configuration, and the mean RISE varies by less than 3\% across
	the grid. The number of retained frequency bins $N_{c}$ falls by a
	factor of up to 32 between the finest ($N_{b}=1$, no coarse-graining) and
	coarsest ($N_{b}=16$, $N_{h}=2$) configurations, with no measurable
	loss of accuracy or calibration. MCMC convergence is satisfactory
	across all cells: $\hat{R}\leq 1.01$ for all parameters, negligible
	divergent transitions, and E-BFMI $>0.6$.

	Wall-clock runtime is essentially constant across the entire grid
	(8.7--9.6\,s on a 4-core CPU node) despite the 32-fold variation in
	$N_{c}$. At this problem size ($K=10$, $p=3$, $n=16{,}384$) the
	per-gradient cost is small enough that runtime is dominated by fixed
	NUTS warm-up, JIT compilation, and SVI initialisation overheads
	rather than by the $\mathcal{O}(p^{2}\,N_{c}\,K)$ likelihood evaluation.
	The benefit of coarse-graining at this scale is therefore primarily
	statistical (a smaller effective $N_{b}N_{h}$ makes the posterior
	easier to sample) rather than computational. Coarse-graining begins
	to pay off in wall-clock terms once $N_{c}$ or $K$ is large enough
	that the per-gradient cost exceeds the fixed overheads, as in the
	LISA application below.

	\begin{table*}[!htbp]
		\centering
		\caption{3D VAR(2) simulation results at $n=16{,}384$, $K=10$,
		degree-2 splines (mean $\pm$ std over $500$ realisations),
		all runs at $\eta=1$. Coverage is the fraction of (frequency,
		matrix-element) pairs inside the pointwise 90\% credible interval.
		RISE is the relative mean squared (Frobenius) error. ESS
		is the median effective sample size across parameters. $N_{h}=$off
		denotes no coarse-graining ($N_{h}=1$).}
		\label{tab:3d_var2_results}
    \centering
    \setlength{\tabcolsep}{10pt}
    \begin{NiceTabular}{ccccccc}
    \CodeBefore
        \rowcolor{gray!10}{2-7}
        \rowcolor{gray!10}{13-16}
        \rowcolor{gray!10}{20-21}
    \Body
        $N_{b}$ & $N_{h}$ & $N_{c}$ & Coverage & RISE & ESS & Runtime (s) \\
		\midrule
		1 & off & 8192 & $0.900 \pm 0.031$ & $0.111 \pm 0.004$ & $21,019 \pm 2726$ & $9.61 \pm 1.76$ \\
		1 & 2 & 4096 & $0.898 \pm 0.031$ & $0.112 \pm 0.005$ & $24,057 \pm 2807$ & $9.21 \pm 1.69$ \\
		1 & 4 & 2048 & $0.900 \pm 0.030$ & $0.112 \pm 0.005$ & $26,352 \pm 3081$ & $9.03 \pm 1.56$ \\
		1 & 8 & 1024 & $0.904 \pm 0.029$ & $0.111 \pm 0.005$ & $26,640 \pm 3303$ & $8.92 \pm 1.65$ \\
		1 & 16 & 512 & $0.908 \pm 0.029$ & $0.110 \pm 0.004$ & $25,480 \pm 3525$ & $8.81 \pm 1.62$ \\
		1 & 32 & 256 & $0.907 \pm 0.030$ & $0.109 \pm 0.004$ & $24,011 \pm 3500$ & $8.82 \pm 1.70$ \\
		2 & off & 4096 & $0.898 \pm 0.031$ & $0.112 \pm 0.005$ & $23,938 \pm 2939$ & $9.11 \pm 1.63$ \\
		2 & 2 & 2048 & $0.901 \pm 0.030$ & $0.112 \pm 0.005$ & $26,279 \pm 2979$ & $9.03 \pm 1.72$ \\
		2 & 4 & 1024 & $0.904 \pm 0.030$ & $0.111 \pm 0.005$ & $26,405 \pm 3361$ & $8.89 \pm 1.67$ \\
		2 & 8 & 512 & $0.907 \pm 0.030$ & $0.110 \pm 0.004$ & $25,726 \pm 3461$ & $8.76 \pm 1.75$ \\
		2 & 16 & 256 & $0.907 \pm 0.031$ & $0.109 \pm 0.004$ & $23,997 \pm 3787$ & $8.81 \pm 1.69$ \\
		4 & off & 2048 & $0.901 \pm 0.030$ & $0.112 \pm 0.004$ & $26,053 \pm 2906$ & $9.05 \pm 1.71$ \\
		4 & 2 & 1024 & $0.904 \pm 0.029$ & $0.111 \pm 0.005$ & $26,695 \pm 3221$ & $8.84 \pm 1.64$ \\
		4 & 4 & 512 & $0.907 \pm 0.030$ & $0.110 \pm 0.004$ & $25,753 \pm 3636$ & $8.75 \pm 1.72$ \\
		4 & 8 & 256 & $0.907 \pm 0.030$ & $0.109 \pm 0.004$ & $24,036 \pm 3622$ & $8.78 \pm 1.68$ \\
		8 & off & 1024 & $0.904 \pm 0.030$ & $0.112 \pm 0.005$ & $26,336 \pm 3190$ & $8.89 \pm 1.65$ \\
		8 & 2 & 512 & $0.907 \pm 0.030$ & $0.111 \pm 0.005$ & $25,587 \pm 3658$ & $8.78 \pm 1.70$ \\
		8 & 4 & 256 & $0.907 \pm 0.031$ & $0.110 \pm 0.004$ & $24,000 \pm 3787$ & $8.79 \pm 1.70$ \\
		16 & off & 512 & $0.905 \pm 0.029$ & $0.111 \pm 0.004$ & $25,504 \pm 3431$ & $8.74 \pm 1.55$ \\
		16 & 2 & 256 & $0.907 \pm 0.030$ & $0.110 \pm 0.004$ & $23,862 \pm 3724$ & $8.84 \pm 1.68$ \\
    \end{NiceTabular}
	\end{table*}

	\subsection{LISA}
	\label{subsec:lisa}
	We apply the multivariate P-spline model to the three second-generation
	time-delay interferometry (TDI) Michelson channels $X_{2}$, $Y_{2}$,
	$Z_{2}$ from a publicly released LISA TDI dataset
	\citep{lisa_noise4a, lisa_noise5a}. All channels are TDI 2.0 throughout,
	and we suppress the generation subscript in what follows, writing \XYZ for
	$(X_{2}, Y_{2}, Z_{2})$. The dataset was
	produced with \texttt{lisainstrument}~\citep{lisainstrument} under the
	LISA Data Challenge (LDC) Spritz noise model~\citep{ldcspritz} at a sampling rate of
	$f_{s}=0.5$\,Hz and constant $8.3$\,s light travel times. The six
	inter-spacecraft interferometer (ISI) and test-mass (TM) link
	measurements were propagated through \texttt{pytdi}
	\citep{staab_pytdi_2025} to produce the TDI 2.0 Michelson combinations used
	here.

	The noise-orthogonal $(A, E, T)$ TDI combinations are obtained from \XYZ by
	the orthonormal rotation $(A, E, T)^{\top} = M_{\AET}\,(X, Y, Z)^{\top}$
	with
	\begin{equation}
		\label{eq:M_AET}
		M_{\AET} = \begin{pmatrix}
			-1/\sqrt{2} & 0 & 1/\sqrt{2} \\
			1/\sqrt{6} & -2/\sqrt{6} & 1/\sqrt{6} \\
			1/\sqrt{3} & 1/\sqrt{3} & 1/\sqrt{3}
		\end{pmatrix},
	\end{equation}
	which satisfies $M_{\AET}^{\top}M_{\AET} = I$ and has unit
	determinant, so the rotation introduces no log-likelihood Jacobian
	correction~\cite{prince2002_LISA_optimal_sensitivity}. Under the
	idealised equal-arm, equal-noise assumption the \AET basis diagonalises
	the noise covariance, motivating the diagonal-\AET restrictions tested
	below.

	We analyse two noise configurations:
	\begin{itemize}
		\item \texttt{noise4a}~\cite{lisa_noise4a} --- symmetric realisation: all six MOSAs
		are assigned identical Optical Metrology System (OMS; ISI carrier) and test-mass acceleration
		amplitude spectral densities equal to the LDC baseline values
		($2.4\times10^{-15}$\,m\,s$^{-2}$/$\sqrt{\mathrm{Hz}}$ and
		$7.9\times10^{-12}$\,m/$\sqrt{\mathrm{Hz}}$, respectively). In this
		regime the $3\times3$ cross-spectral matrix is close to degenerate:
		$S_{XX}\!\approx\!S_{YY}\!\approx\!S_{ZZ}$, with off-diagonals equal in
		magnitude. A closed-form reference is available from the
		analytic TDI transfer of the link-level ASDs.
		\item \texttt{noise5a}~\cite{lisa_noise5a} --- asymmetric realisation: the six
		per-MOSA OMS and test-mass ASDs are drawn independently from
		$\mathcal{U}(0.5,\,2.0)\times$baseline with a fixed seed. The
		resulting cross-spectral matrix has distinct diagonals and
		non-trivial off-diagonal structure, representing the physically
		relevant regime where mis-modelled cross-spectral correlations
		would degrade sensitivity to stochastic backgrounds. A
		closed-form $3\times3$ reference is built with the
		\texttt{SEGWO} package~\citep{bayle_segwo} by feeding the
		twelve measured per-MOSA ASDs through analytic OMS/TM filter
		shapes and projecting through PyTDI's TDI 2.0 Michelson
		\XYZ combinations to obtain the full
		$3\times3$ \XYZ covariance at every frequency.
	\end{itemize}
	We fit each dataset under three competing hypotheses for the
	$3\times3$ noise cross-spectrum, designed to test whether the
	\XYZ$\to$\AET rotation diagonalises the LISA noise covariance to
	within statistical resolution. Since $M_{\AET}$ is
	orthogonal with unit determinant, fitting in the \AET basis is
	equivalent to fitting in the \XYZ basis up to a deterministic
	relabelling of $\S$. Wherever the \AET basis is used
	($\mathcal{H}_{0}$ and $\mathcal{H}_{1}$), the \XYZ$\to$\AET
	rotation is applied to the time series first, and all subsequent
	blocking, averaging, and coarse-graining is performed in the
	fitting basis. The three hypotheses therefore form a
	strictly nested hierarchy over the families of representable
	spectral matrices,
	$\mathcal{H}_{0}\subset\mathcal{H}_{1}\subset\mathcal{H}_{2}$.
	The three models also differ in their data object:
	$\mathcal{H}_{0}$ uses the three diagonal \AET periodograms,
	whereas $\mathcal{H}_{1}$ and $\mathcal{H}_{2}$ use the full
	$3\times3$ periodogram matrix (in the \AET and \XYZ bases
	respectively):
	\begin{itemize}
		\item $\mathcal{H}_{0}$ (\emph{diagonal \AET}).
		The cross-spectral matrix is constrained to be \emph{exactly}
		diagonal in the \AET basis. The \XYZ time series is first
		rotated into \AET via $M_{\AET}$ (Eq.~\ref{eq:M_AET}),
		and each of the three \AET channels is then fit independently
		with a univariate ($p=1$) log-P-spline model on the diagonals
		$\log\delta_{j}^{2}$ ($j=1,2,3$, corresponding to
		$\log S_{AA}, \log S_{EE}, \log S_{TT}$) with $K=100$ basis
		functions each. Only the log-diagonal expansion
		Eq.~\eqref{eq:delta_spline} is used. The off-diagonal
		expansions Eqs.~\eqref{eq:theta_re_spline}
		and~\eqref{eq:theta_im_spline} are not invoked.
		The data for this model are the three \emph{univariate}
		averaged periodograms --- the diagonal entries
		$S_{AA}, S_{EE}, S_{TT}$ of $\Y_{h}^{\mathrm{cg}}$ --- so the
		likelihood is a product of three independent Gamma
		(univariate Whittle) likelihoods, one per \AET channel; the
		off-diagonal periodogram entries never enter the fit. This is
		precisely the special case of the complex Wishart likelihood
		with a \emph{diagonal} scale matrix $\S$, which factorises
		into the three diagonal Gamma terms. The diagonal \AET posterior is assembled
		per-sample as $\mathrm{diag}(S_{AA},S_{EE},S_{TT})$ and
		rotated back to the \XYZ basis via
		$\S_{\XYZ}=M_{\AET}^{\top}\S_{\AET}M_{\AET}$
		for diagnostics. The \XYZ off-diagonals carried by
		$\mathcal{H}_{0}$ therefore reflect only the deterministic
		TDI rotation, with no fitted physical cross-power.
		\item $\mathcal{H}_{1}$ (\emph{multivariate \AET with
		restricted off-diagonals}). The cross-spectral matrix is
		\emph{approximately} diagonal in the \AET basis. Unlike
		$\mathcal{H}_{0}$, the data here are the full $3\times3$
		averaged \AET periodogram matrix and the likelihood is the
		complete multivariate complex Wishart --- the off-diagonal
		entries do enter the fit. What is restricted is the
		off-diagonal \emph{model}, not the data. The full
		multivariate Cholesky model
		(Eqs.~\ref{eq:delta_spline}--\ref{eq:theta_im_spline})
		is fit to the \AET-rotated time series with $K=100$ basis
		functions on the diagonals $\log\delta_{j}^{2}$ but only $K_{\theta}=2$
		basis functions on each off-diagonal component $\theta_{jl}$, forcing
		any fitted cross-power to be smooth and small.
		$\mathcal{H}_{1}$ thus occupies the natural middle ground
		between the strictly diagonal $\mathcal{H}_{0}$ (off-diagonal
		data discarded, no cross-power) and the unrestricted
		$\mathcal{H}_{2}$: it sees the off-diagonal data but is
		biased toward a near-diagonal solution, so it can reveal
		smooth cross-power when present without introducing
		spurious structure when it is not. The posterior spectral matrix is
		rotated back to the \XYZ basis for diagnostics.
		\item $\mathcal{H}_{2}$ (\emph{full multivariate \XYZ}).
		The cross-spectral matrix has fully populated off-diagonals.
		The multivariate Cholesky model
		(Eqs.~\ref{eq:delta_spline}--\ref{eq:theta_im_spline})
		is fit directly to the \XYZ time series with $K=100$ basis
		functions for \emph{every} Cholesky component, both diagonal and
		off-diagonal. No rotation is
		needed since the fit is already in the \XYZ basis. This is
		the unrestricted reference model.
	\end{itemize}
	For \texttt{noise4a} the \AET basis diagonalises the noise
	exactly, so the strong diagonal restriction imposed by
	$\mathcal{H}_{0}$ (and the milder one of $\mathcal{H}_{1}$)
	are both expected to be valid descriptions, and all three
	models should agree within statistical resolution. For
	\texttt{noise5a} the per-MOSA asymmetry introduces genuine
	off-diagonal power in the \AET basis: $\mathcal{H}_{0}$ is
	then mis-specified by construction (no off-diagonal flexibility
	at all), $\mathcal{H}_{1}$ can absorb only the small-bandwidth
	component of the off-diagonals through its $K_{\theta}=2$
	basis functions, and $\mathcal{H}_{2}$ remains unrestricted. The expected
	fit-quality ordering is therefore
	$\mathrm{RISE}(\mathcal{H}_{0})\!>\!\mathrm{RISE}(\mathcal{H}_{1})\!>\!\mathrm{RISE}(\mathcal{H}_{2})$
	for \texttt{noise5a} and
	$\mathrm{RISE}(\mathcal{H}_{0})\!\approx\!\mathrm{RISE}(\mathcal{H}_{1})\!\approx\!\mathrm{RISE}(\mathcal{H}_{2})$
	for \texttt{noise4a}.

	For all three models, knots are placed by the adaptive quantile rule
	described in Section~\ref{sec:psplines} so that more knots fall near
	the steep low-frequency rise. We choose $N_{b}$ to give blocks of
	order one week in length, and a fixed safe-Bayes tempering
	$\eta=0.5$ (justified by the sensitivity sweep in
	Appendix~\ref{app:eta_sweep}). Each block is tapered with a Tukey window
	(taper fraction $0.1$, $\mathrm{ENBW}\approx1.04$) before
	computing the DFT to suppress spectral leakage. The three TDI
	transfer-function nulls at $0.030$, $0.060$, and $0.090$\,Hz are
	excised with half-width $1$\,mHz prior to fitting. All other settings
	follow Section~\ref{sec:bayesian-inference}.

	In addition to the RISE introduced in Section~\ref{subsec:sim_3d_var2},
	we report the median relative width of the 90\% credible band on the
	diagonal PSDs,
	\begin{equation}
		\label{eq:delta_psd}
		\Delta_{\mathrm{PSD}} = \operatorname{median}_{f}
		\frac{\hat{S}^{95\%}(f)-\hat{S}^{5\%}(f)}{\hat{S}^{50\%}(f)},
	\end{equation}
	computed in the \XYZ basis for all three models. This complements the RISE
	(a bias metric) with a summary of posterior precision.
	Table~\ref{tab:lisa_results} reports RISE, $\Delta_{\mathrm{PSD}}$, and
	wall-clock duration for the three models, two datasets, and three
	observation durations. All eighteen LISA runs satisfy the convergence
	diagnostics of
	Section~\ref{sec:bayesian-inference}: $\hat R<1.01$ for all parameters,
	bulk and tail ESS exceeding $400$, E-BFMI~$>0.3$, and negligible
	divergent transitions with no tree-depth saturation. The results split sharply
	along the \AET-diagonalisation hypothesis.

	\emph{Symmetric noise (\texttt{noise4a}).} All three models recover
	the analytic spectral matrix at the half-percent level
	($\mathrm{RISE}\approx 4.4\!-\!5.3\times10^{-3}$) across all
	durations and are statistically indistinguishable from one another.
	This is the expected behaviour when the \AET basis truly diagonalises
	the noise: $\mathcal{H}_{0}$ already captures the cross-spectral
	structure because the only off-diagonal power present in the \XYZ
	basis is the deterministic geometric coupling recovered by the
	$M_{\AET}^{\top}\!\cdot\!M_{\AET}$ rotation, and
	the additional off-diagonal flexibility of $\mathcal{H}_{1}$ and
	$\mathcal{H}_{2}$ has nothing to fit. Two conclusions follow:
	(i) for equal-arm LISA noise the multivariate analysis is not
	required, and three independent univariate fits suffice, and
	(ii) the unrestricted $\mathcal{H}_{2}$ model does not lose
	accuracy on data that does not need its flexibility, confirming
	that $\mathcal{H}_{2}$ remains a safe default when the
	diagonal-\AET assumption cannot be guaranteed in advance.

	\emph{Asymmetric noise (\texttt{noise5a}).} The three models
	separate by orders of magnitude in the order predicted by their
	off-diagonal flexibility,
	$\mathrm{RISE}(\mathcal{H}_{0})\!>\!\mathrm{RISE}(\mathcal{H}_{1})\!>\!\mathrm{RISE}(\mathcal{H}_{2})$.
	$\mathcal{H}_{0}$ stalls at
	$\mathrm{RISE}\approx 3.3\!\times\!10^{-2}$ at all durations ---
	a floor set by the \AET off-diagonal power it is structurally
	unable to represent. $\mathcal{H}_{1}$ improves on this by
	roughly $50\%$
	($\mathrm{RISE}\approx 1.8\!-\!2.2\!\times\!10^{-2}$) by
	allowing two off-diagonal basis functions, but is still limited
	by its narrow $\theta_{jk}$ basis. $\mathcal{H}_{2}$ reaches
	$\mathrm{RISE}\approx 9\!-\!11\!\times\!10^{-4}$, about
	$20\times$ smaller than $\mathcal{H}_{1}$ and $30\!-\!40\times$
	smaller than $\mathcal{H}_{0}$, and continues to improve with
	duration. This monotone nested ordering is the central diagnostic
	of the LISA study: when the per-MOSA noise levels are asymmetric
	(the physically realistic regime) the \AET rotation does
	\emph{not} diagonalise the cross-spectral matrix, and a
	multivariate model that estimates the full Cholesky factor is
	required.

	Posterior uncertainty contracts as expected. For the multivariate
	models, the median relative CI width on the diagonal PSDs falls
	from $\approx\!4.7\%$ at 1\,month to $\approx\!1.2\%$ at 1\,year,
	a reduction of roughly $4\times$, consistent with the $\sqrt{T}$
	scaling expected for variance-dominated Whittle inference, and is
	essentially identical between $\mathcal{H}_{1}$ and
	$\mathcal{H}_{2}$ and across datasets, confirming that the
	per-channel resolution is set by the data and not by the choice
	of basis. The univariate $\mathcal{H}_{0}$ contracts at the same
	$\sqrt{T}$ rate but with consistently narrower bands
	($\approx\!3.4\%$ at 1\,month, $\approx\!0.9\%$ at 1\,year),
	reflecting the smaller per-channel parameter count of the $p=1$
	model when off-diagonals are removed from the inference. For
	\texttt{noise5a} this tighter $\mathcal{H}_{0}$ posterior is a
	misleading indicator of fit quality --- the bands do not include
	the truth because the model is structurally unable to represent
	the \AET off-diagonals, as the order-of-magnitude RISE gap to
	$\mathcal{H}_{1}$ and $\mathcal{H}_{2}$ makes clear.

	\begin{table*}[!htbp]
		\centering
		\caption{LISA TDI \XYZ results at fixed safe-Bayes
			tempering $\eta=0.5$ for the symmetric (\texttt{noise4a}) and
			asymmetric (\texttt{noise5a}) datasets, fit with the
			diagonal-\AET null model ($\mathcal{H}_{0}$,
			off-diagonal Cholesky coefficients fixed to zero), the \AET-rotated multivariate
			model with restricted off-diagonal basis ($\mathcal{H}_{1}$,
			$K_{\theta}=2$), and the unrestricted multivariate \XYZ
			model ($\mathcal{H}_{2}$), at three observation
			durations. $N_{b}$ is the number of Wishart
			time blocks (chosen to give blocks of approximately
			seven days in length) and $N_{c}$ is set to
			$1024$. RISE is the relative mean squared (Frobenius)
			error of the posterior-median spectral matrix
			against the analytic reference (closed-form equal-link
			LDC for \texttt{noise4a}, SEGWO with measured per-MOSA
			ASDs for \texttt{noise5a}), computed in the \XYZ basis
			for all three models (the \AET-basis models are first
			rotated back to \XYZ via $M_{\AET}$).
			$\Delta_{\rm PSD}$ is the median over frequencies of
			the pointwise relative width
			$[\hat{S}^{95\%}(f)-\hat{S}^{5\%}(f)]/\hat{S}^{50\%}(f)$
			of the 90\% credible band on the diagonal PSDs (also
			in the \XYZ basis). Runtime is wall-clock seconds for
			one full SVI+NUTS pipeline run on a single 4-core CPU
			node; for $\mathcal{H}_{0}$ this is the total over the
			three independent univariate channel fits.}
		\label{tab:lisa_results}
		\setlength{\tabcolsep}{5pt}
		\begin{NiceTabular}{llc ccc ccc ccc}
		\CodeBefore
			\rowcolor{gray!10}{3-5}
		\Body
			& & & \multicolumn{3}{c}{$\mathcal{H}_{0}$} & \multicolumn{3}{c}{$\mathcal{H}_{1}$} & \multicolumn{3}{c}{$\mathcal{H}_{2}$} \\
			\cmidrule(lr){4-6}\cmidrule(lr){7-9}\cmidrule(lr){10-12}
			Noise & Duration & $N_{b}$ & RISE & $\Delta_{\rm PSD}$ & $t$\,(s) & RISE & $\Delta_{\rm PSD}$ & $t$\,(s) & RISE & $\Delta_{\rm PSD}$ & $t$\,(s) \\
			\midrule
			\multirow{3}{*}{\texttt{4a}}
				& $1$\,mo & $4$  & $4.3\!\times\!10^{-3}$ & $3.4\%$ & $16$  & $4.5\!\times\!10^{-3}$ & $4.8\%$ & $117$ & $4.7\!\times\!10^{-3}$ & $4.6\%$ & $119$ \\
				& $6$\,mo & $25$ & $4.7\!\times\!10^{-3}$ & $1.3\%$ & $13$  & $4.7\!\times\!10^{-3}$ & $1.9\%$ & $224$ & $5.3\!\times\!10^{-3}$ & $1.9\%$ & $181$ \\
				& $1$\,yr & $52$ & $4.4\!\times\!10^{-3}$ & $0.8\%$ & $244$ & $4.4\!\times\!10^{-3}$ & $1.2\%$ & $632$ & $5.0\!\times\!10^{-3}$ & $1.2\%$ & $853$ \\
			\multirow{3}{*}{\texttt{5a}}
				& $1$\,mo & $4$  & $3.3\!\times\!10^{-2}$ & $3.5\%$ & $15$  & $2.2\!\times\!10^{-2}$ & $4.9\%$ & $119$ & $1.1\!\times\!10^{-3}$ & $4.5\%$ & $177$ \\
				& $6$\,mo & $25$ & $3.3\!\times\!10^{-2}$ & $1.4\%$ & $13$  & $1.8\!\times\!10^{-2}$ & $1.9\%$ & $218$ & $9.8\!\times\!10^{-4}$ & $1.9\%$ & $245$ \\
				& $1$\,yr & $52$ & $3.3\!\times\!10^{-2}$ & $1.0\%$ & $191$ & $2.2\!\times\!10^{-2}$ & $1.3\%$ & $728$ & $8.6\!\times\!10^{-4}$ & $1.3\%$ & $1089$ \\
		\end{NiceTabular}
	\end{table*}

	\begin{figure*}[!t]
		\centering
		\includegraphics[width=0.48\linewidth]{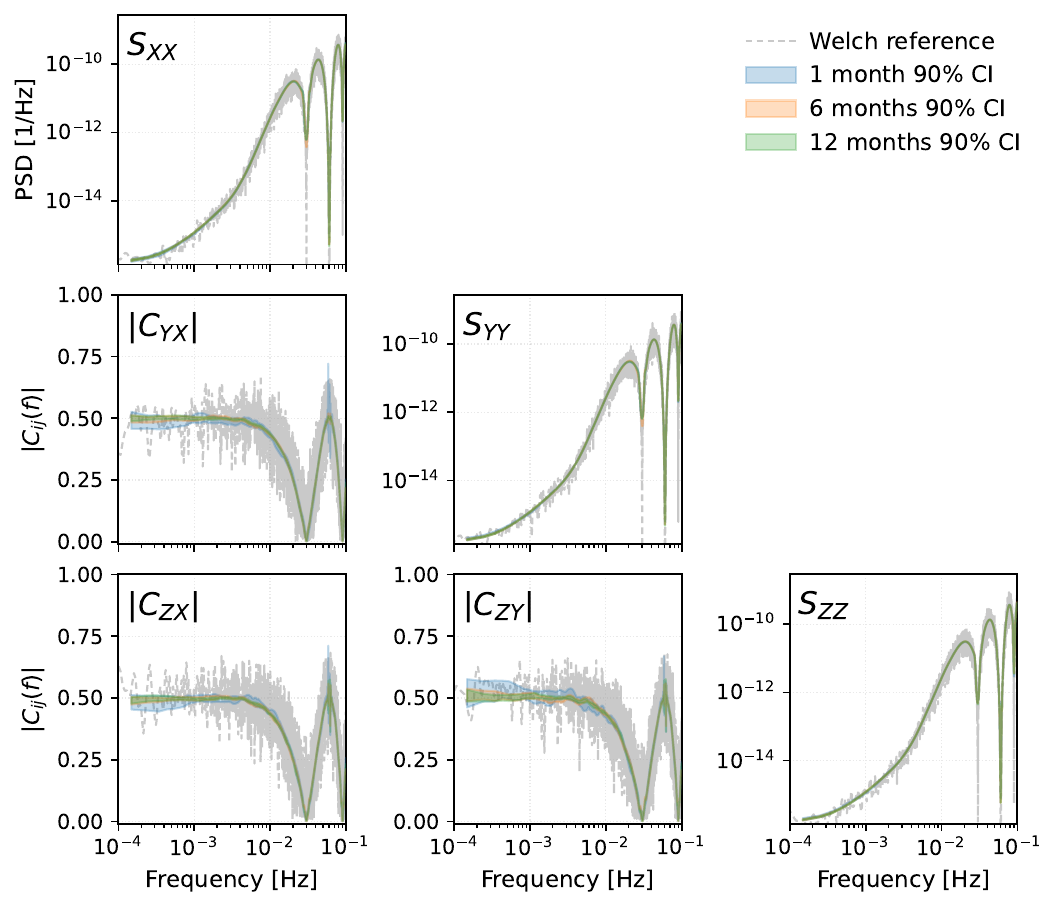}\hfill
		\includegraphics[width=0.48\linewidth]{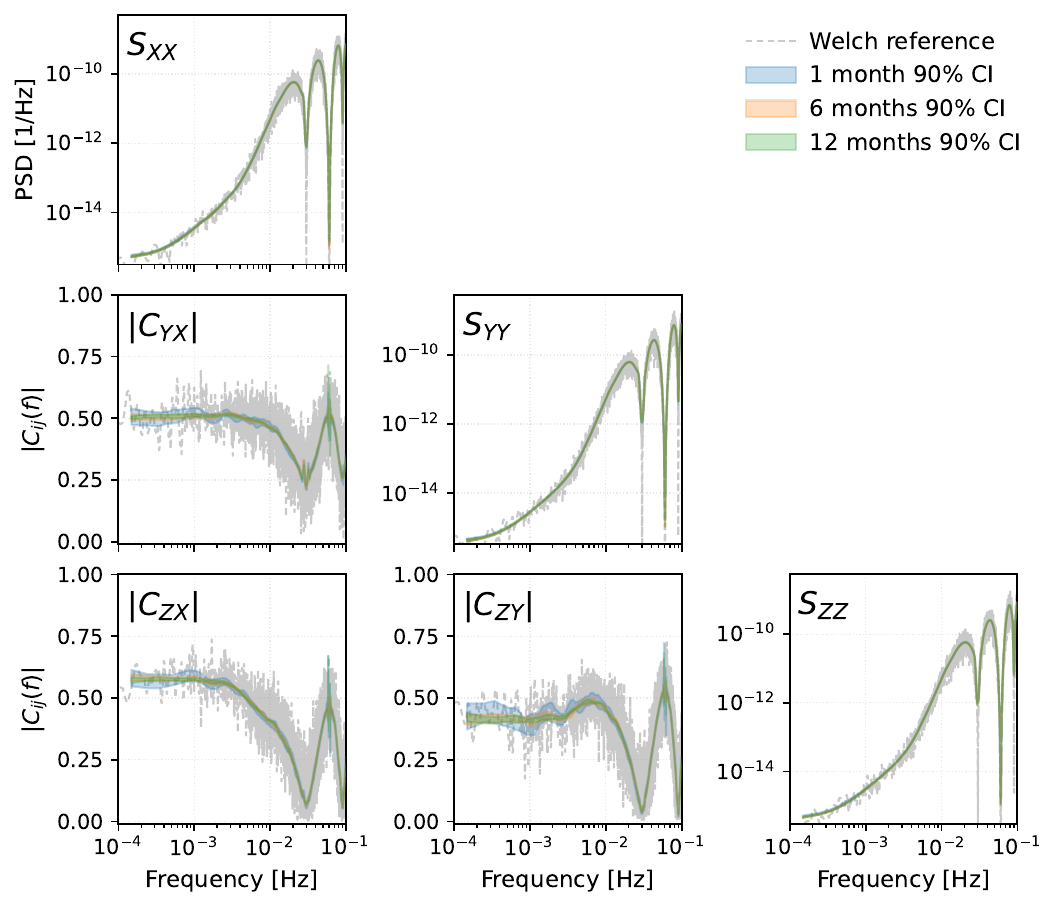}
		\caption{Posterior summaries of the LISA TDI \XYZ
			spectral matrix from the unrestricted
			$\mathcal{H}_{2}$ model for the symmetric \texttt{noise4a}
			(left) and asymmetric \texttt{noise5a} (right) datasets at
			safe-Bayes tempering $\eta=0.5$, shown at three observation
			durations: $1$ month (blue), $6$ months (orange), and $12$
			months (green). Diagonal panels show the posterior $90\%$
			credible bands for the diagonal PSDs
			$S_{XX},S_{YY},S_{ZZ}$. Lower-triangle panels show the
			coherence $|C_{ij}(f)|$ (Eq.~\ref{eq:coherence}) for each pair. Welch reference
			(grey) is overplotted.}
		\label{fig:lisa_eta0p5}
		\script{plot_lisa_triangle.py}
	\end{figure*}

	Figure~\ref{fig:lisa_eta0p5} overlays the three durations for both
	datasets under the $\mathcal{H}_{2}$ model. The
	diagonal PSDs are recovered with sub-percent bias and CI widths that
	contract with $\sqrt{T}$, while the off-diagonal coherences (the
	cross-channel quantities most directly tied to
	stochastic-background and parameter-estimation
	pipelines~\citep{katz2024globalfit}) are resolved at the
	few-percent level by 6 months and tightened by a further factor of
	$\sim 1.7$ at 1 year. A localised feature visible in all three
	off-diagonal panels of Fig.~\ref{fig:lisa_eta0p5} is a cluster of
	small-amplitude wiggles in the posterior-median coherence near the
	TDI transfer-function null at $\approx 0.06$\,Hz (and, more weakly,
	near $0.03$ and $0.09$\,Hz). These arise because the diagonal PSDs
	$S_{ii}(f)$ drop by several orders of magnitude across each null,
	and the P-spline basis --- which is smooth on a log scale but
	agnostic to the location and shape of the dips --- has to absorb
	this dynamic range from the data alone. Small residual misfits at
	the bottoms of the dips are amplified in the coherence, producing
	the visible ripple. Incorporating the analytic TDI transfer
	functions directly into the prior mean (rather than only excising
	a $1$\,mHz window around each null) may be a natural way to
	suppress this artefact and is left for future work.

	\section{Discussion}
	\label{sec:discussion}

	We have presented a multivariate P-spline framework for estimating the
	cross-spectral density matrix of stationary vector time series. The
	Cholesky parametrisation of $\S^{-1}$ guarantees Hermitian positive
	definiteness at every frequency and factorises the likelihood into $p$
	independent per-channel regressions that can be sampled in parallel.
	A blocked, coarse-grained Whittle likelihood combined with safe-Bayes
	$\eta$-tempering yields near-nominal 90\% credible-interval coverage
	($\approx 0.90$) across a wide $(N_{b},N_{h})$ grid on the 3D VAR(2)
	benchmark, while reducing the number of inference-time frequency bins by
	up to a factor of $32$.

	On the synthetic 3D VAR(2) benchmark the un-tempered estimator
	($\eta=1$) achieves coverage $\approx\!0.90$ at every $(N_{b},N_{h})$
	configuration tested. On the LISA TDI data the same un-tempered
	estimator over-concentrates. A fixed safe-Bayes tempering $\eta=0.5$
	recovers credible bands that contract with $\sqrt{T}$ and remain
	well-behaved across all three observation durations
	(Appendix~\ref{app:eta_sweep}).

	\paragraph{$\eta$-tempering and debiased Whittle.}
	The $\eta$-sweep of Appendix~\ref{app:eta_sweep} shows that
	credible-band width is a strong, monotone function of $\eta$ at fixed
	$(N_{b},N_{h},K)$, but does not yet identify a principled rule for
	choosing $\eta$ across regimes. Mapping the optimal $\eta$ as a
	function of $N_{b}N_{h}$, $K$, and the underlying spectral curvature
	is the most pressing methodological question raised by this work.
	A complementary direction is to replace $\eta$-tempering with a
	debiased Whittle likelihood~\citep{sykulski2019_debiased}, which
	corrects the finite-sample bias of the periodogram analytically
	rather than down-weighting the entire likelihood.

	\paragraph{Informative priors.}
	The hierarchical Gaussian smoothing prior used here is
	near-uninformative: each weight has marginal variance set by a
	high-level hyperparameter ($\phi_{j}, \nu_{j}$) with no scale
	imposed by the data ahead of inference. In practice this places no
	bound on where the spline curves can sit, and most of NUTS
	warm-up is spent locating a region of weight space consistent with
	even a coarse Welch estimate. A more informative prior, centred on
	a fast initial estimate of $\log\delta_{j}^{2}(f)$ and the
	off-diagonal Cholesky terms (e.g.\ from a Welch-and-smooth pre-pass),
	with $\phi_{j}$ controlling allowed deviations from that estimate,
	may (i) make the smoothing hyperparameters physically meaningful as
	fractional excursion bounds, (ii) shorten warm-up substantially, and
	(iii) reduce the prior--data contention that drives over-coverage at
	small $\eta$ in the appendix sweep.

	\paragraph{Non-stationarity.}
	The Whittle likelihood assumes wide-sense stationarity. Both
	\texttt{noise4a} and \texttt{noise5a} are stationary by
	construction: realisations of an LDC noise model with constant
	per-MOSA amplitude spectral densities and constant $8.3$\,s light
	travel times. The real LISA noise environment is non-stationary on
	multiple timescales --- constellation breathing modulates the arm
	lengths over the year-long orbital period, and per-link noise levels
	drift on timescales from hours (thermal cycling, glitches) to
	months (component aging). A natural and tractable extension is to
	assume \emph{piecewise} stationarity: partition the mission timeline
	into segments short enough that within-segment stationarity is a
	good approximation but long enough that the Whittle approximation
	remains accurate, and either fit each segment independently or
	couple consecutive segments through a smoothness prior on the
	frequency-dependent Cholesky parameters in the time direction. This
	would extend the present model to a time-frequency spectral matrix
	$\S(t,f)$ at the cost of one additional smoothing dimension.

	\paragraph{Improved variational families.}
	As reported in Section~\ref{subsec:sim_3d_var2}, the current low-rank
	($r=16$) multivariate Gaussian SVI guide recovers an accurate
	posterior median but produces narrower credible bands than NUTS,
	with a corresponding drop in empirical coverage. NUTS is therefore
	retained as the production sampler, and SVI is used only as an
	initialiser. Because the model is implemented in NumPyro, the
	variational family is a swappable component: replacing the Gaussian
	guide with a normalising flow (e.g.\ a neural spline flow or a
	masked autoregressive / inverse autoregressive flow) is essentially
	a one-line change. A systematic study of which variational families
	recover calibrated PSD posteriors --- so that SVI could serve as a
	stand-alone approximation at a fraction of the NUTS wall-clock cost
	--- is left for future work.

	\paragraph{Other extensions.}
	The within-bin constant-$\S$ approximation underlying coarse-graining
	(Eq.~\ref{eq:log-like-coarse}) can bias estimates near sharp
	spectral features. Adaptive bin widths that narrow around lines and
	transfer-function nulls are a natural extension. Hybrid
	parametric/nonparametric models that embed known instrumental
	features within the P-spline envelope are another route to improving
	sharp-feature accuracy without sacrificing the smoothness elsewhere.

	\subsection*{Data and Software Availability}
	The software developed for this project is open-source and publicly available from the GitHub repository
	\url{https://github.com/nz-gravity/LogPSplinePSD.git}, which contains all source code, example scripts,
	and configuration files needed to reproduce the results.
	Installation instructions and dependency information (tested on Python~$\geq 3.10$) are provided in the repository \texttt{README}.
	The in-text code links (\scalebox{0.65}{\GitHubIcon}) are pinned to the tagged release \texttt{v0.1.0} of the package, so that the referenced line numbers remain valid as the codebase evolves.
	The software is released under the MIT License, permitting free use, modification, and distribution.
	The LISA time series data are available at~\citep{lisa_noise4a, lisa_noise5a}.
    Scripts for the LISA analysis are available from the GitHub repository
	\url{https://github.com/nz-gravity/lisa_multivar_logpsplines_study.git}.

	\begin{acknowledgments}
		We thank Quentin Baghi, Jean-Baptiste Bayle, Ollie Burke, Nikolaos
		Karnesis, Martina Muratore, Federico Pozzoli, and others in the LISA Noise Non-Stationarities
		Group (part of the ``Deep analysis group'' of the Distributed Data Processing
		Centre, DDPC) for helpful discussions. We additionally thank Jean-Baptiste
		Bayle for providing the LISA dataset, and the University of Glasgow for the computing
		resources that supported JB in simulating the dataset. We
		acknowledge the use of the \texttt{SEGWO}
		package~\citep{bayle_segwo}, developed by J.-B.~Bayle and
		O.~Hartwig, used here to construct the analytic $3\times3$
		cross-spectral reference for the asymmetric \texttt{noise5a}
		realisation. PMR, RM, and AV
		gratefully acknowledge support from the Marsden Fund Council grants MFP-UOA2131 and MFP-UOA2531,
		funded by the New Zealand Government and managed by the Royal Society Te Apārangi.
		This work was performed on the OzSTAR national facility at Swinburne
		University of Technology. The OzSTAR program receives funding in part from
		the Astronomy National Collaborative Research Infrastructure Strategy (NCRIS)
		allocation provided by the Australian Government, and from the Victorian Higher
		Education State Investment Fund (VHESIF) provided by the Victorian Government.
	\end{acknowledgments}

	\appendix

	\section{Notation}
	\label{app:notation}

	Table~\ref{tab:definitions} lists the symbols used throughout the
	paper, grouped by the stage of the model in which they appear.

	\input{definitions_table}

	\section{Bivariate VAR(2) benchmark}
	\label{app:var2_appendix}

	This appendix benchmarks the proposed estimator against the bivariate
	Bayesian baselines available in the literature on a controlled VAR(2)
	problem with closed-form ground truth.
	Following~\citet{yixuan_vnpc}, we simulate 500 independent realisations of a bivariate
	VAR(2) time series at three sample sizes $n=256,512,1024$
	(refer to~\citet[Section~4.2,][]{yixuan_vnpc} for the
	definitions of the VAR(2) models), and the theoretical spectral
	density matrix provides exact ground truth. The spectral densities are estimated using
	the multivariate P-spline method with $K=16$ basis functions and degree-2 splines
	($d=2$) for each spectral component. We compare to the variational Bayes (VB) and variational
	nonparametric correction (VNPC) methods of~\citet{yixuan_vnpc} using the same
	$L_{2}$ error criterion. The $L_{2}$ error is the un-normalised
	integrated Frobenius error of the posterior-median spectral matrix,
	without dividing by $\int\lVert S(f)\rVert_{F}\,\mathrm{d}f$. We use
	it here for direct comparison with the VB and VNPC results, which are
	reported in this metric.

	Table~\ref{tab:var2_l2_results} summarises the $L_{2}$ errors across the 500
	independent realisations. The multivariate P-spline estimator is competitive
	with both comparators at every sample size and attains the lowest mean error in
	all three settings. The mean $L_{2}$ error decreases from $0.095$ at $n=256$ to
	$0.072$ at $n=512$ and $0.055$ at $n=1024$, compared with $0.121$, $0.091$, and
	$0.066$ for VB and $0.129$, $0.103$, and $0.082$ for VNPC. This is the main
	message we want from the bivariate benchmark: the proposed multivariate P-spline method
	achieves accuracy of the same order as established alternatives, while in these
	simulations being modestly more accurate on average.

	\begin{table}[!t]
    \centering
    \caption{Bivariate VAR(2) benchmark using the $L_{2}$ error metric from
   ~\citet{yixuan_vnpc}. Entries are mean $\pm$ standard deviation over 500
    independent realisations. Smaller values are better.}
    \label{tab:var2_l2_results}
    \footnotesize
    \setlength{\tabcolsep}{4pt}
    \begin{NiceTabular}{@{}lccc@{}}
        \toprule
        {$n$} & {P-spline} & {VB} & {VNPC} \\
        \midrule
        256 & $0.095 \pm 0.022$ & $0.121 \pm 0.029$ & $0.129 \pm 0.033$ \\
        512 & $0.072 \pm 0.018$ & $0.091 \pm 0.027$ & $0.103 \pm 0.025$ \\
        1024 & $0.055 \pm 0.013$ & $0.066 \pm 0.016$ & $0.082 \pm 0.017$ \\
        \bottomrule
		\end{NiceTabular}
	\end{table}

	\section{$\eta$-tempering sensitivity on \texttt{noise4a}}
	\label{app:eta_sweep}

\begin{figure}[!t]
		\centering
		\includegraphics[width=\linewidth]{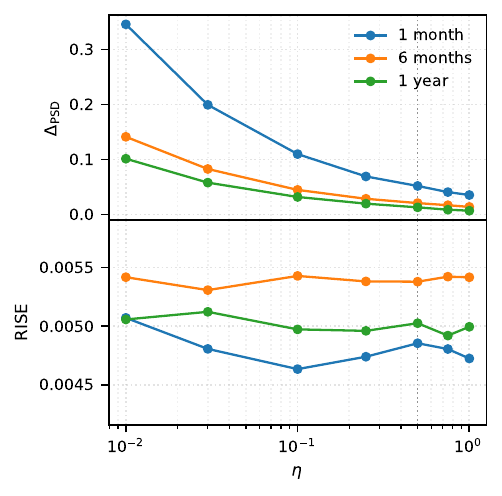}
		\caption{Safe-Bayes $\eta$-sensitivity on the symmetric LISA
			\texttt{noise4a} dataset at three observation durations
			(1\,month, 6\,months, 1\,year).
			Top: median pointwise relative width of the 90\% credible
			band on the diagonal PSDs.
			Bottom: matrix RISE (narrow $y$-range---RISE varies by
			${\lesssim}5\%$ across the full sweep).
			The vertical dotted line marks $\eta=0.5$, the value used in
			Section~\ref{subsec:lisa}.}
		\label{fig:lisa_eta_sweep}
		\script{plot_lisa_eta.py}
	\end{figure}

	To support the choice of tempering parameter for the LISA application,
	we re-ran the \texttt{noise4a} dataset at $\eta\in\{0.01, 0.03, 0.1,
	0.25, 0.5, 0.75, 1\}$ for each of the three observation durations
	(1, 6, and 12 months) used in Section~\ref{subsec:lisa}, holding all
	other settings fixed. Figure~\ref{fig:lisa_eta_sweep} summarises the
	result.

	Two features emerge from Figure~\ref{fig:lisa_eta_sweep}.
	First, the matrix RISE is essentially flat across the entire sweep:
	it varies by less than $\pm5\%$ about its mean at every duration,
	confirming that the posterior-median point estimate is nearly
	insensitive to $\eta$.
	Second, credible-band width contracts monotonically with $\eta$, by
	roughly an order of magnitude from $\eta=0.01$ to $\eta=1$, with
	the steepest contraction at the longest duration---consistent with
	the per-bin Fisher information scaling as $N_{b}N_{h}$
	(Eq.~\eqref{eq:fisher-scaling}).

	The vertical dotted line marks the choice $\eta=0.5$ used in
	Section~\ref{subsec:lisa}. This value gives a practical
	compromise: credible bands are meaningfully wide at all three
	durations while the point estimate remains accurate throughout.
	Reaching nominal coverage would require yet smaller $\eta$ or a
	more flexible model (e.g.\ adaptive coarse-graining near the TDI
	nulls). We leave this to future work.

	\bibliography{bib}
\end{document}

%% file: definitions_table.tex
\begin{table*}[t]
    \centering
    \footnotesize
    \begingroup
    \setlength{\fboxsep}{6pt}%
    \newcommand{\defitem}[2]{%
        \parbox[t]{0.18\textwidth}{\raggedright #1}\hfill
        \parbox[t]{0.78\textwidth}{\raggedright #2}\par\smallskip
    }
    \newcommand{\groupplain}[2]{%
        \par\addvspace{2pt}%
        \noindent\parbox{\textwidth}{\textit{#1}\par\smallskip #2}\par
        \addvspace{2pt}%
    }
    \newcommand{\groupshaded}[2]{%
        \par\addvspace{2pt}%
        \noindent\colorbox{gray!8}{%
            \parbox{\dimexpr\textwidth-2\fboxsep\relax}{%
                \textit{#1}\par\smallskip #2%
            }%
        }\par
        \addvspace{2pt}%
    }
    \noindent\hrulefill\par\smallskip
    \noindent\parbox[t]{0.18\textwidth}{\textbf{Symbol}}\hfill
    \parbox[t]{0.78\textwidth}{\textbf{Meaning}}\par\smallskip
    \hrulefill\par\smallskip

    \groupshaded{Time series \& sampling}{%
        \defitem{$T,\ \Delta_t,\ f_s$}{Total observation duration, sampling interval, sampling frequency; $T=n\Delta_t$, $\Delta_t=1/f_s$.}
        \defitem{$f_{Ny}$}{Nyquist frequency, $f_{Ny}=f_s/2$.}
        \defitem{$n,\ p$}{Number of time samples per channel, number of channels.}
        \defitem{$\Delta_f,\ f_k$}{Frequency resolution $\Delta_f=1/T$; $k$th DFT frequency $f_k=k\Delta_f$.}
        \defitem{$N$}{Number of positive Fourier frequencies, $N=\lfloor n/2\rfloor$.}
    }

    \groupplain{Blocking \& coarse-graining}{%
        \defitem{$N_b,\ T_b$}{Number of non-overlapping time blocks; block duration $T_b=T/N_b$ ($n/N_b$ samples per block).}
        \defitem{$N_h,\ J_h,\ \bar{f}_h$}{Number of fine frequencies aggregated per coarse bin; $J_h$ the $h$th bin (consecutive subset of fine frequencies), with midpoint $\bar f_h$.}
        \defitem{$N_c$}{Number of coarse-grain bins, $N_c=n/(2N_bN_h)$.}
    }

    \groupshaded{Spectral density}{%
        \defitem{$\Z$}{Observed $p$-channel time series, $\Z\in\mathbb{R}^{n\times p}$.}
        \defitem{$\d(f_k)$}{DFT vector, $\d(f_k)\in\mathbb{C}^p$.}
        \defitem{$\Gamma(\tau),\ \gamma_{lm}(\tau)$}{Autocovariance matrix at integer lag $\tau$ and its $(l,m)$ entry.}
        \defitem{$\S(f_k)$}{$p\times p$ Hermitian positive-definite spectral density matrix; Fourier transform of $\Gamma$.}
    }

    \groupplain{Periodogram \& Wishart statistics}{%
        \defitem{$\I^{(i)}(f_k),\ \bar\I(f_k)$}{Block-$i$ periodogram, $\I^{(i)}=\d^{(i)}\d^{(i)*}\sim\mathcal{CW}_p(T_b\S,1)$; block-averaged periodogram $\bar\I=N_b^{-1}\sum_i\I^{(i)}$.}
        \defitem{$\Y(f_k)$}{Summed periodogram / Wishart statistic, $\Y=N_b\bar\I\sim\mathcal{CW}_p(T_b\S,N_b)$.}
        \defitem{$\Y_h^{\mathrm{cg}}$}{Coarse-grained statistic, $\Y_h^{\mathrm{cg}}=\sum_{f_k\in J_h}\Y(f_k)\,\dot\sim\,\mathcal{CW}_p(T_b\S(\bar f_h),N_bN_h)$.}
        \defitem{$\lambda_\nu^{(k)},\ \v_\nu^{(k)},\ \u_\nu^{(k)}$}{$\nu$th eigenvalue/unit eigenvector of $\Y(f_k)$; scaled eigenvector $\u_\nu^{(k)}=\sqrt{\lambda_\nu^{(k)}}\,\v_\nu^{(k)}$.}
    }

    \groupshaded{Cholesky parametrisation}{%
        \defitem{$\mathbf{T}_k,\ \mathbf{D}_k$}{Factors in $\S(f_k)^{-1}=\mathbf{T}_k^*\mathbf{D}_k^{-1}\mathbf{T}_k$: $\mathbf{T}_k$ unit lower-triangular, $\mathbf{D}_k$ diagonal with entries $\delta_{1k}^2,\ldots,\delta_{pk}^2$.}
        \defitem{$\theta_{jl}^{(k)},\ \delta_{jk}^2$}{Off-diagonal entry $(j,l)$ of $\mathbf{T}_k$ ($j>l$) and $j$th diagonal entry of $\mathbf{D}_k$.}
        \defitem{$\btheta_j,\ \bdelta_j$}{Parameter vectors for channel $j$: collect $\theta_{jl}^{(h)}$ ($l<j$) and $\delta_{jh}$ across the coarse bins $h=1,\ldots,N_c$.}
    }

    \groupplain{P-spline model}{%
        \defitem{$K,\ B_m(f_k)$}{Number of B-spline basis functions (model hyperparameter, distinct from frequency index $k$); $m$th basis evaluated at $f_k$.}
        \defitem{$K_j,\ K_{jl}$}{Basis sizes for the $j$th diagonal and $(j,l)$ off-diagonal components.}
        \defitem{$\mathbf{w}_j^{(\delta)},\ \mathbf{w}_{jl}^{(\Re)},\ \mathbf{w}_{jl}^{(\Im)}$}{Spline coefficient vectors for $\log\delta_{jk}^2$, $\Re[\theta_{jl}]$, $\Im[\theta_{jl}]$.}
        \defitem{$\mathbf{P}_j,\ \phi_j$}{Penalty matrix $[\mathbf{P}_j]_{lm}=\int_0^1 B_l''B_m''\,\mathrm{d}t$ and smoothing precision hyperparameter.}
    }

    \par\smallskip\hrulefill
    \endgroup
    \caption{Notation used throughout the paper. Symbols are grouped by
        the stage of the model they belong to.}
    \label{tab:definitions}
\end{table*}